\documentclass[structabstract]{aa}  

\usepackage{graphicx}
\usepackage{amsmath} 
\usepackage[varg]{txfonts}
\usepackage{pdflscape}

\usepackage{natbib,twoopt}
\usepackage{hyperref} 
\bibpunct{(}{)}{;}{a}{}{,} 
\makeatletter
\newcommandtwoopt{\citeads}[3][][]{\href{http://adsabs.harvard.edu/abs/#3}%
{\def\hyper@linkstart##1##2{}%
\let\hyper@linkend\@empty\citealp[#1][#2]{#3}}}
\newcommandtwoopt{\citepads}[3][][]{\href{http://adsabs.harvard.edu/abs/#3}%
{\def\hyper@linkstart##1##2{}%
\let\hyper@linkend\@empty\citep[#1][#2]{#3}}}
\newcommandtwoopt{\citetads}[3][][]{\href{http://adsabs.harvard.edu/abs/#3}%
{\def\hyper@linkstart##1##2{}%
\let\hyper@linkend\@empty\citet[#1][#2]{#3}}}
\newcommandtwoopt{\citeyearads}[3][][]%
{\href{http://adsabs.harvard.edu/abs/#3}
{\def\hyper@linkstart##1##2{}%
\let\hyper@linkend\@empty\citeyear[#1][#2]{#3}}}
\makeatother

\usepackage{color}
\definecolor{mygreen}{RGB}{0,128,0}

\hypersetup{colorlinks=true,linkcolor=blue,citecolor=blue,urlcolor=blue}

\usepackage{etoolbox}
\makeatletter
\patchcmd\@combinedblfloats{\box\@outputbox}{\unvbox\@outputbox}{}{%
   \errmessage{\noexpand\@combinedblfloats could not be patched}%
}%
 \makeatother

\newcommand*{\ditto}{--~\raisebox{-2pt}{\texttt{"}}~--}

\begin{document} 

   \title{The \textit{Gaia} reference frame for bright sources\\ examined using VLBI observations of radio stars}

   \subtitle{}

   \author{Lennart Lindegren
          }

   \institute{Lund Observatory, Department of Astronomy and Theoretical Physics, Lund University,
   Box 43, SE-22100 Lund, Sweden\\
              \email{lennart@astro.lu.se}
             }

   \date{ }

 
\abstract
  {Positions and proper motions of \textit{Gaia} sources are expressed in a reference frame 
  that ideally should be non-rotating relative to distant extragalactic objects, coincident with the 
  International Celestial Reference System (ICRS), and consistent across all magnitudes. 
  For sources fainter than 16th magnitude, this is achieved through \textit{Gaia}'s 
  direct observations of quasars. At brighter magnitudes, it is difficult to validate the 
  quality of the reference frame because comparison data are scarce.}  
  {The aim of this paper is to examine the use of very long baseline interferometry (VLBI) 
  observations of radio stars to determine 
  the spin and orientation of the bright reference frame of current and future \textit{Gaia} data releases.}  
  {Simultaneous estimation of the six spin and orientation parameters makes optimal use of VLBI data 
  and makes it possible to include even single-epoch VLBI observations in the solution. The method 
  is applied to \textit{Gaia} Data Release 2 (DR2) using published VLBI data for 41 radio stars.}  
  {The VLBI data for the best-fitting 26 sources indicate that the bright reference frame 
  of \textit{Gaia} DR2 rotates relative to the faint quasars at a rate of about 0.1~mas~yr$^{-1}$, which is
  significant at the $2\sigma$ level. This supports a similar 
  conclusion based on a comparison with stellar positions in the \textsc{Hipparcos} frame. The accuracy 
  is currently limited because only a few radio sources are included in the solution, by uncertainties in 
  the \textit{Gaia} DR2 proper motions, and by issues related to the astrophysical nature of the radio stars.}  
  {While the origin of the indicated rotation is understood and can be avoided in future data releases, 
  it remains important to validate the bright reference frame of \textit{Gaia} by independent 
  observations. This can be achieved using VLBI astrometry, which may require re-observing 
  the old sample of radio stars as well as measuring new objects. The unique historical value of positional 
  measurements is stressed and VLBI observers are urged to ensure that relevant positional information 
  is preserved for the future.}

   \keywords{astrometry --
                proper motions --
                reference systems --
                instrumentation: interferometers --
                methods: data analysis
               }

   \titlerunning{The \textit{Gaia} reference frame for bright sources} 
   \authorrunning{L.~Lindegren}

   \maketitle

%

\noindent
\textbf{Note:} Owing to a coding error in the implementation of the analysis 
method described in the paper, the results from its application to 
\textit{Gaia} DR2 data, as presented in the original version of the paper 
\citep{2020A&A...633A...1L}, were significantly wrong. This affected 
Tables~2 and 3, Figs.~3--5, and parts of Sects.~3.3, 3.4, 4, and 5. 
A \textit{Corrigendum} based on the corrected code is in press
(\href{https://doi.org/10.1051/0004-6361/201936161e}%
{doi:10.1051/0004-6361/201936161e}).
The present arXiv version (v5) contains the complete paper, including revised tables, 
figures, and portions of the text as published in the \textit{Corrigendum}.

\section{Introduction} 
\label{sec:intro}

The \textit{Gaia} Celestial Reference Frame \citep[\textit{Gaia}-CRF;][]{2018A&A...616A..14G}
is formally defined by the positions, as measured by \textit{Gaia}, of a large number of sources that are
identified as quasars. Through their 
cosmological distances, these objects define a kinematically non-rotating reference frame,
that is, their proper motions are assumed to be zero on average. A subset of them, identified as 
the optical counterparts of radio sources with accurate positions in the International Celestial 
Reference Frame \citep[ICRF;][]{1998AJ....116..516M} from very long baseline interferometry 
(VLBI) observations, are used to align the axes of the 
non-rotating quasar frame with the ICRF. The second release of \textit{Gaia} data 
\citep[DR2;][]{2018A&A...616A...1G} lists 556\,869 quasars whose positions at epoch J2015.5
define the optical reference frame known as \textit{Gaia}-CRF2. This includes 2820 sources 
matched to a prototype version of ICRF3 \citep{2018cosp...42E1583J}.

The vast majority of sources in \textit{Gaia} DR2 are Galactic stars with sizeable proper motions.
The implicit assumption is that the positions and proper motions of the stars, and indeed the
barycentric coordinates of all \textit{Gaia} sources, are expressed in the same reference frame 
as the quasars. This is fundamental for the dynamical interpretation of the observations, which 
assumes the absence of the inertial (Coriolis and centrifugal) forces that would appear in a 
rotating frame.

Although the measurement and reduction principles of the \textit{Gaia} mission have been
designed to provide a globally consistent reference frame for all kinds of objects,
subtle differences are inevitable as a consequence of the varying conditions under 
which the objects are observed. For example, the quasars defining \textit{Gaia}-CRF are all
faint (fewer than 1\% have $G<17$~mag), on average bluer than stars of comparable 
magnitude, and they have a very different distribution on the sky than the stars.
Differences in magnitude, colour, and numerous other factors are likely to produce small shifts of 
the image centroids, which, if left uncalibrated, may propagate into systematic errors of the positions 
and proper motions. 

Intricate instrument models have been set up and calibrated as part of the global astrometric 
reductions of \textit{Gaia} data in order to eliminate such systematics. In \textit{Gaia} DR2 
there is nevertheless an indication that the reference frame defined by the bright stars 
(up to $G\simeq 11$ to 13) rotates with respect to the quasars at a rate of a few tenths of a 
milliarcsecond (mas) per year. This is illustrated in Fig.~4 of \citet{2018A&A...616A...2L} and 
further quantified by 
\citet{2018ApJS..239...31B}. In Brandt's paper the rotation shows up as a systematic offset 
of the proper motions of bright stars in \textit{Gaia} DR2 with respect to the 
``\textsc{Hipparcos}--\textit{Gaia} proper motions'' calculated from the position differences 
between \textit{Gaia} DR2 (at epoch 2015.5) and the \textsc{Hipparcos} catalogue 
(at epoch 1991.25), scaled by the epoch difference of $\sim$24~yr. Through the long time-baseline, the \textsc{Hipparcos}--\textit{Gaia} proper motions constitute a precise set of 
reference values, which are moreover inertial because the positional systems of \textsc{Hipparcos} 
and \textit{Gaia} were both aligned with the International Celestial Reference System (ICRS) 
at their respective epoch. 
The observed offset therefore points to a systematic error in the \textit{Gaia} DR2 proper 
motions of the bright stars, equivalent to an inertial rotation of its reference frame.

The cause of this rotation is discussed in Appendix~\ref{sec:phys}. Briefly, it is related 
to the different modes in which \textit{Gaia}'s CCDs are operated, depending on the magnitude 
of the source, and corresponding differences in the calibration models. In particular, around
$G=13,$ there is a transition in the on-board CCD sampling scheme from two- to 
one-dimensional pixel windows, causing abrupt changes in the quality of both the astrometric 
data and the $G$-band photometry (which uses the same CCDs). Examples of this are shown 
in Figs.~9 and B.2 of \citet{2018A&A...616A...2L}, and Fig.~9 of \citet{2018A&A...616A...4E}. 

It is therefore justified to study the reference frames separately that are defined by the bright and 
the faint \textit{Gaia} sources, and to
draw the division line at $G\simeq 13$. \textit{Gaia}-CRF2, being defined by the quasars, 
clearly belongs to the faint part and is by construction to very high accuracy non-rotating 
with respect to the ICRS; its properties are discussed elsewhere \citep{2018A&A...616A..14G}.
The subject of this paper is the bright reference frame of \textit{Gaia}, defined by the
system proper motions for sources with $G\lesssim13$. Available data, including the 
\textsc{Hipparcos}--\textit{Gaia} proper motions mentioned above, are not sufficient to 
decide if the transition from the faint to the bright reference frame occurs abruptly at 
this magnitude, or more gradually over a few magnitudes; for the purpose of this paper, 
I will in general assume that the transition is abrupt so that all sources brighter than 
$G=13.0$ are in the same reference frame.

Future \textit{Gaia} data releases may provide proper motions that are an order of magnitude more
precise than they were in DR2, and similar or even greater improvements could be obtained in the systematics. 
The quasars, of which many more will be found, will continue to be the main tool for examining the
quality of the \textit{Gaia} reference frame at faint magnitudes. Assessing its consistency at 
brighter magnitudes will be much harder. The \textsc{Hipparcos}--\textit{Gaia} proper 
motions will then be of little use because their random and systematic errors are already dominated 
by the position errors in the \textsc{Hipparcos} celestial reference frame (HCRF), 
which remain unchanged. The quality
of the bright reference frame can only be verified by means of the positions and proper motions 
of bright \textit{Gaia} sources, measured to sufficient accuracy in the ICRF frame by some 
independent method. Most obviously, this can be done by means of differential VLBI, where
the positions of radio stars are measured relative to quasars by phase-referencing techniques 
\citep{1990AJ.....99.1663L,1995ASPC...82..327B,2011AJ....141..114R,2012aamm.book..175F}. 
These relative measurements are already reaching microarcsecond ($\mu$as) precision \citep{2014ARA&A..52..339R}. 

The purpose of this paper is to explore the use of differential VLBI observations of radio stars
for verification of \textit{Gaia}'s bright reference frame. In Sect.~\ref{sec:theory} the required
formalism is developed, whereby the positions and proper motions measured by VLBI are
connected to
the \textit{Gaia} data. The method is tested on \textit{Gaia} DR2 data in Sect.~\ref{sec:demo},
using a selection of published VLBI observations, and the results and possible future improvements 
are discussed in Sect.~\ref{sec:disc}.

\section{Theory}
\label{sec:theory}

\subsection{ICRS and celestial reference frames}
\label{sec:ICRS}

The ICRS is an idealised system of astronomical
coordinates $\alpha$, $\delta$, whose axes are defined by convention and remain fixed with 
respect to distant matter in the Universe \citep{1995A&A...303..604A}. The origin is at the 
Solar System barycentre.
Any astrometric catalogue where the positions and proper motions nominally refer to the ICRS 
can be regarded as a practical realisation of the idealised system, and is then called a 
celestial reference frame (CRF). The ICRF, HCRF, and \textit{Gaia}-CRF2 are examples of such 
reference frames, among which the ICRF has the privileged status of actually defining the 
conventional axes of the ICRS. In the present context it is necessary to consider that
\textit{Gaia} DR2 may represent (at least) two distinct reference frames, one defined 
by the faint quasars, and a second defined by the positions and proper motion
of the bright stars in \textit{Gaia} DR2. 

\begin{figure}[t]
\centerline{
\includegraphics[scale=0.30,clip=true]{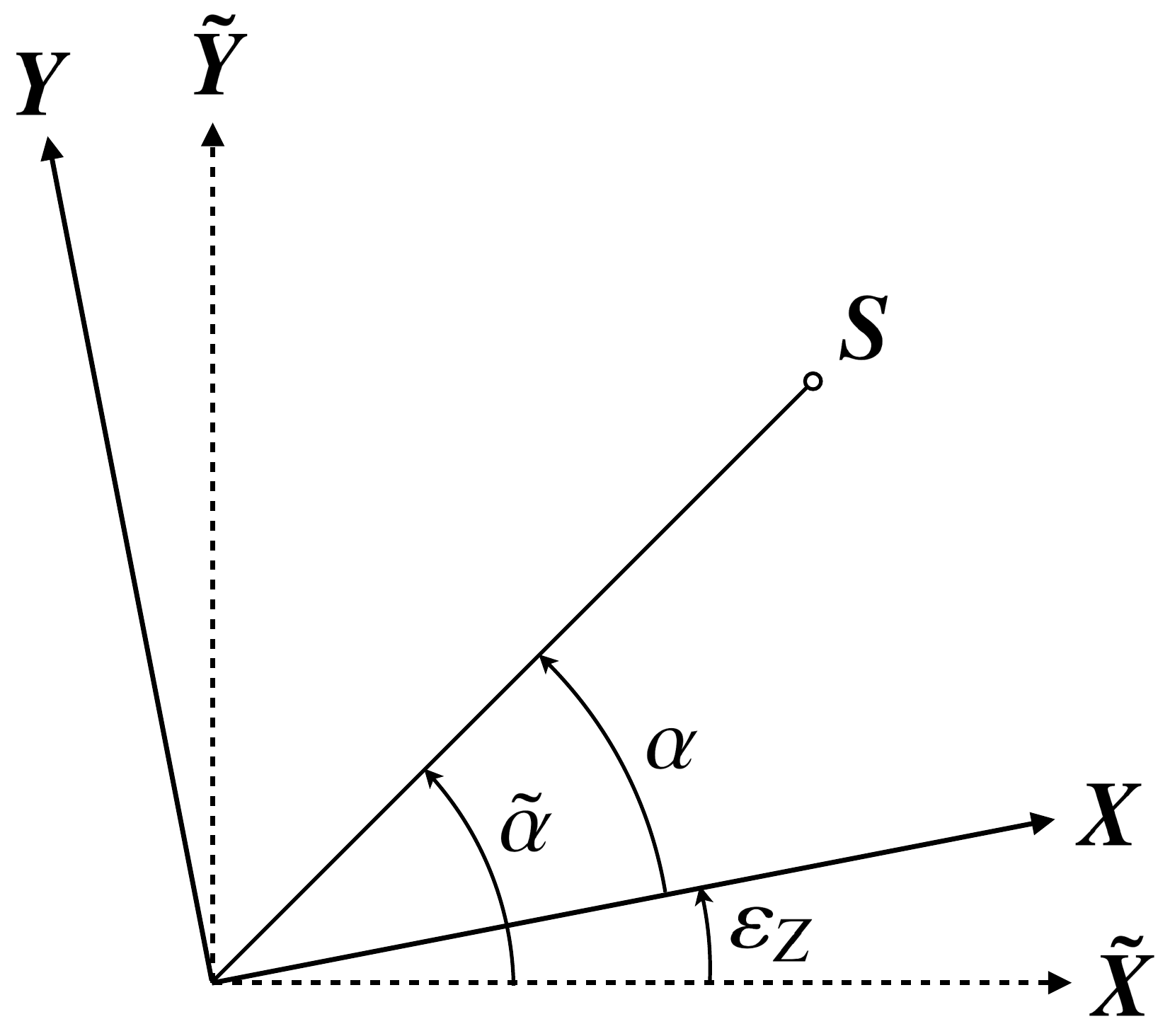}
}
\caption{Convention for the definition of the orientation of the arbitrary frame 
$\tens{\tilde{C}}=[\vec{\tilde{X}}~\vec{\tilde{Y}}~\vec{\tilde{Z}}]$ with 
respect to the ICRF ($\tens{C}=[\vec{X}~\vec{Y}~\vec{Z}]$). 
The drawing illustrates the configuration when the orientation difference is a pure rotation about 
the $Z$ axes by the positive angle $\varepsilon_Z$, i.e.\ $\vec{\varepsilon}=[0,\,0,\,\varepsilon_Z]'$.
The right ascension of the source at $\vec{S}$ is $\alpha$ in frame $\tens{C}$ and 
$\tilde{\alpha}=\alpha+\varepsilon_Z$ in frame $\tens{\tilde{C}}$.\label{fig01}}
\end{figure}

Conceptually, a CRF can be visualised as a set of orthogonal unit vectors $\vec{X}$, $\vec{Y}$,
$\vec{Z}$, with origin at the Solar System barycentre, and with $\vec{X}$ 
pointing towards $\alpha=\delta=0$, $\vec{Z}$ towards $\delta=+90^\circ$, and 
$\vec{Y}=\vec{Z}\times\vec{X}$. Let $\tens{C}=[\vec{X}~\vec{Y}~\vec{Z}]$ be the 
vector triad representing the 
ICRF. By definition, $\tens{C}$ coincides with the axes of the ICRS and is fixed with respect to
objects at cosmological distances such as the quasars. This means that the proper motions of
quasars, when expressed in $\tens{C}$, have no global component that can be interpreted
as a solid-body rotation (spin) of $\tens{C}$.    
Any other reference frame $\tens{\tilde{C}}=[\vec{\tilde{X}}~\vec{\tilde{Y}}~\vec{\tilde{Z}}]$ 
may have some small time-dependent 
offset from $\tens{C}$ described by the vector $\vec{\varepsilon}(t)$,
\begin{equation}\label{e04}
\tens{C} = \tens{\tilde{C}} + \vec{\varepsilon}(t) \times \tens{\tilde{C}} + O(\varepsilon^2)\, .
\end{equation}
Thus $\vec{\varepsilon}(t)$ is the rotation of $\tens{\tilde{C}}$ needed to align its axes
with $\tens{C}$. The sign of $\vec{\varepsilon}(t)$ is chosen for consistency with earlier publications 
\citep{1995A&A...304..189L, agis2012, 2016A&A...595A...4L}, 
where the frame offset was defined in the sense of a correction to the frame under investigation. 
The components of $\vec{\varepsilon}(t)$ in $\tens{C}$ or $\tens{\tilde{C}}$ are denoted 
$\varepsilon_X(t)$, $\varepsilon_Y(t)$, $\varepsilon_Z(t)$. The relation between the two frames
is illustrated in Fig.~\ref{fig01}.

Equation~(\ref{e04}) is valid in the small-angle approximation, that is,\ ignoring terms of order 
$\varepsilon^2$, where $\varepsilon=|\vec{\varepsilon}|$ is the total angular offset between 
the two frames. This is a valid approximation in all practical cases, where $\varepsilon\lesssim 1$~mas,
or $\varepsilon^2<5\times 10^{-9}$~mas (5~pas). Rigorous expressions are given in Section~6.1.2 of 
\citet{agis2012}.

The relation between the two frames can alternatively be expressed by means of the rotation matrix
\begin{equation}\label{eq:rotMat}
\tens{C}'\tens{\tilde{C}}=\begin{bmatrix}
\vec{X}'\vec{\tilde{X}} & \vec{X}'\vec{\tilde{Y}} & \vec{X}'\vec{\tilde{Z}} \\ 
\vec{Y}'\vec{\tilde{X}} & \vec{Y}'\vec{\tilde{Y}} & \vec{Y}'\vec{\tilde{Z}} \\ 
\vec{Z}'\vec{\tilde{X}} & \vec{Z}'\vec{\tilde{Y}} & \vec{Z}'\vec{\tilde{Z}} \end{bmatrix}
= \begin{bmatrix} 1 & \varepsilon_Z & -\varepsilon_Y\\
-\varepsilon_Z & 1 & \varepsilon_X\\ \varepsilon_Y & -\varepsilon_X & 1 \end{bmatrix} 
+ O(\varepsilon^2) \, ,
\end{equation}
where the prime ($'$) denotes matrix transposition and the scalar product of vectors
\citep{murray1983}. 
Let $\vec{u}$ be the unit vector at a certain time from the Solar System barycentre towards a 
celestial object. The rectangular coordinates of the vector in the two frames are given by
\begin{equation}\label{eq:Cu}
\tens{C}'\vec{u}=\begin{bmatrix} \cos\alpha\cos\delta\\ \sin\alpha\cos\delta\\ \sin\delta\end{bmatrix}
\, ,\quad
\tens{\tilde{C}}'\vec{u}=\begin{bmatrix} \cos\tilde{\alpha}\cos\tilde{\delta}\\ 
\sin\tilde{\alpha}\cos\tilde{\delta}\\ \sin\tilde{\delta}\end{bmatrix}\, ,
\end{equation}
where $(\alpha,~\delta)$ and $(\tilde{\alpha},~\tilde{\delta})$ are the astronomical coordinates
of the object in the two frames. The column matrices in Eq.~(\ref{eq:Cu}) are related by the matrix equation
\begin{equation}\label{eq:uu}
\tens{C}'\vec{u} = (\tens{C}'\tens{\tilde{C}})\,\tens{\tilde{C}}'\vec{u} \, ,
\end{equation}
where $\tens{C}'\tens{\tilde{C}}$ is the rotation matrix in Eq.~(\ref{eq:rotMat}). To verify Eq.~(\ref{eq:uu}),
one can note that $\tens{\tilde{C}}\tens{\tilde{C}}'$ is the unit tensor \citep{murray1983}, and hence
$\tens{\tilde{C}}\tens{\tilde{C}}'\vec{u}=\vec{u}$.

In frame $\tens{\tilde{C}}$ the proper motions are modelled as essentially constant angular velocities, 
which does not permit $\tens{\tilde{C}}$ to have a non-uniform rotation with respect to distant matter. 
The variation of the offset vector with time $t$ can therefore be written
\begin{equation}\label{e08}
\vec{\varepsilon}(t) = \vec{\varepsilon}(T) + (t-T)\,\vec{\omega} \,  ,
\end{equation}    
where $T$ is an arbitrary reference epoch, and $\vec{\varepsilon}(T)$ and $\vec{\omega}$ are
constant vectors with components $\varepsilon_X(T)$, $\omega_X$, etc.

Equation~(\ref{e08}) describes a uniform solid rotation of one frame relative to the other.
The word ``rotation'' is ambiguous in this context because it may refer to either the instantaneous
configuration $\vec{\varepsilon}(t)$ or the angular velocity $\vec{\omega}$.
In the following we use ``spin'' for the angular velocity and ``orientation'' for the instantaneous 
configuration, specifically $\vec{\varepsilon}(T)$; for brevity, the combined or general effect 
may be called ``rotation''.

\subsection{Differences in position and proper motion}
\label{sec:diff}

From Eqs.~(\ref{eq:rotMat})--(\ref{eq:uu}), the following first-order expressions are obtained for the 
coordinate differences:
\begin{align}\label{e06}
(\alpha-\tilde{\alpha})\cos\delta &= 
+\varepsilon_X\cos\alpha\sin\delta+\varepsilon_Y\sin\alpha\sin\delta-\varepsilon_Z\cos\delta \, ,\\
\delta-\tilde{\delta} &= 
-\varepsilon_X\sin\alpha\phantom{\cos\delta\,}+\varepsilon_Y\cos\alpha \, .\label{e07}
\end{align}
The time derivative of the above gives the corresponding expressions for the proper motion differences,
\begin{align}\label{e01}
\mu_{\alpha*}-\tilde{\mu}_{\alpha*} &= 
+\omega_X\cos\alpha\sin\delta+\omega _Y\sin\alpha\sin\delta-\omega_Z\cos\delta \, ,\\
\mu_{\delta}-\tilde{\mu}_{\delta} &= 
-\omega_X\sin\alpha\phantom{\cos\delta\,}+\omega_Y\cos\alpha \, ,\label{e02}
\end{align}
where 
$\mu_{\alpha*}=(\text{d}\alpha/\text{d}t)\cos\delta$,
$\mu_{\delta}=\text{d}\delta/\text{d}t$ 
are the components of the proper motion in frame $\tens{C}$ and 
$\tilde{\mu}_{\alpha*}=(\text{d}\tilde{\alpha}/\text{d}t)\cos\tilde{\delta}$,
$\tilde{\mu}_{\delta}=\text{d}\tilde{\delta}/\text{d}t$
the components in $\tens{\tilde{C}}$. In the small-angle approximation, one can use either 
set of coordinates, $(\alpha,\delta)$ or $(\tilde{\alpha},\tilde{\delta})$, for the trigonometric 
factors in (\ref{e06})--(\ref{e02}); the choice made here is arbitrary. 

The use of equations such as Eqs.~(\ref{e06})--(\ref{e02}) for estimating the difference in orientation 
and spin between two astrometric catalogues has been well established in the literature for many years
\citep[among many others, e.g.,][]{1977VeARI..28....1F,1982A&A...116...89F,1988A&A...199..357A,
1991A&A...245..669B,1995A&A...304..182L,2000PASP..112.1103Z,2004A&A...413..771M,
2011MNRAS.416..403F,2015AstL...41..156B}. To estimate the difference in spin ($\vec{\omega}$),
the typical procedure has been to set up Eqs.~(\ref{e01})--(\ref{e02}), using the proper motion 
differences for a number of sources with accurate proper motions in both catalogues (or for 
which the true proper motions can be assumed to be negligible), and solve the resulting 
overdetermined system of equations using the least-squares method. It is thus possible to
estimate $\vec{\omega}$ without knowing $\vec{\varepsilon}$  because the differences in 
$\alpha$ and $\delta$ between the two catalogues are of second order in 
Eqs.~(\ref{e01})--(\ref{e02}).

The orientation difference ($\vec{\varepsilon}$) can be estimated by applying a similar procedure 
to the position differences, resulting in a set of equations like Eqs.~(\ref{e06})--(\ref{e07}) for the
three unknowns $\varepsilon_X(T)$, $\varepsilon_Y(T)$, and $\varepsilon_Z(T)$. If the sources have
proper motion, a complication is that the position differences must be computed for a fixed 
common epoch ($T$), which may require one or both sets of positions to be propagated from 
their mean epoch of observation. This propagation must in addition take into account any difference 
in spin between the two catalogues. Except when both catalogues contain only sources 
with zero proper motion, it is therefore usually not possible to estimate $\vec{\varepsilon}(T)$ independently 
of $\vec{\omega}$.

The general procedure should consequently consider the joint
estimation of $\vec{\varepsilon}(T)$ and $\vec{\omega}$. This may in fact lead to a much
better determination of $\vec{\omega}$ than if only proper motion differences are used. 
The details of the procedure are worked out below, but the basic idea is simple enough: 
if a set of independent positional differences are obtained for a range of epochs,
the resulting equations (\ref{e06})--(\ref{e07}) will depend on both $\vec{\varepsilon}(T)$ 
and $\vec{\omega}$, allowing all six parameters to be determined. In the current context, 
this means that positional VLBI observations of \textit{Gaia} sources, suitably spread out in time,
will contribute to the determination of the spin. This is true even when there is only a single
epoch of VLBI data per source, so that their proper motions (and parallaxes) cannot be 
determined purely from the VLBI observations.

The realisation that positional observations contribute to the determination of the spin is 
of course not new. It was implicit in several of the methods used to link the \textsc{Hipparcos}
catalogue to the ICRS \citep{1997A&A...323..620K}, and explicitly discussed by 
\citet[][Ch.~7.4]{Walter+Sovers2000}, who concluded that it might become desirable 
to revise the \textsc{Hipparcos} link if, in the future, many more radio stars obtained 
accurate interferometric positions.

\subsection{Joint estimation of the rotation parameters}
\label{sec:est}

The joint estimation of $\vec{\varepsilon}(T)$ and $\vec{\omega}$ from 
the \textit{Gaia} and VLBI data for a certain set $S$ of common sources is now considered. It is assumed that the
\textit{Gaia} observations refer to frame $\tens{\tilde{C}}$ and the VLBI observations to
$\tens{C}$ ($=\text{ICRS}$), with Eqs.~(\ref{e04}) and (\ref{e08}) connecting the frames. 
For conciseness, the six unknown rotation parameters are written as the column matrix
$\vec{x}=[\varepsilon_X(T),\varepsilon_Y(T),\varepsilon_Z(T),\omega_X,\omega_Y,\omega_Z]'$.
The result is an estimate of $\vec{x}$, denoted $\vec{\hat{x}}$, together with its $6\times 6$ 
covariance matrix.
As long as the full covariance of $\vec{\hat{x}}$ is retained, the choice of $T$ is in principle arbitrary, 
and for convenience, we adopt the same reference epoch as for the \textit{Gaia} data, 
for example $T=\text{J2015.5}$ for \textit{Gaia} DR2.

The six rotation parameters in $\vec{x}$ are not the only unknowns of the problem. Both the 
VLBI and the \textit{Gaia} observations provide information on the astrometric parameters of
the sources in the common set $S$, and a basic assumption is that each source $i$ in $S$ can 
only have one set of 
``true'' astrometric parameters, here denoted by the column matrix $\vec{y}_i$. Considering
$m=|S|$ sources, the end result of the estimation process consists of $\vec{\hat{x}}$ and 
the $m$ estimates $\vec{\hat{y}}_i$ for $i=1\dots m$. Effectively, each $\vec{\hat{y}}_i$ is the 
weighted mean of the astrometric parameters as determined by VLBI and by \textit{Gaia}, 
after correcting the latter values for the estimated frame rotation $\vec{\hat{x}}$.

As discussed in Sect.~\ref{sec:out}, the comparison of \textit{Gaia} and VLBI measurements
is potentially affected by a number of difficulties including radio--optical offsets and non-linear
motions. In order to proceed with the theoretical development, these difficulties are 
ignored here. It is assumed that the optical and radio data refer to the same physical point 
source, and that this source moves through space at uniform velocity relative to the Solar System 
barycentre. Astrometrically, then, the radio star is completely described by the usual five parameters 
$\alpha$, $\delta$, $\varpi$ (parallax), $\mu_{\alpha*}$, and $\mu_\delta$, referred to the adopted 
epoch $T$, and the radial velocity $v_r$, assumed to be known from spectroscopy. These 
parameters describe the ``true'' motion of the source in frame $\tens{C}$, and differ 
in general both from the \textit{Gaia} parameters and from those derived from the VLBI 
measurements. The subsequent treatment is vastly simplified if expressions are linearised 
around a fixed set of reference values, which is an acceptable approximation as the differences
are typically much smaller than an arcsecond for a suitable choice of reference values (cf.\ the 
discussion in Sect.~\ref{sec:ICRS}).

It is convenient to use the astrometric parameters as given by \textit{Gaia} as reference values 
for the linearisation. The parameter array for source $i$,
\begin{equation}\label{eq:yi}
\vec{y}_i = \left[ \begin{array}{l} \Delta\alpha*_i\\ \Delta\delta_i\\ \Delta\varpi_i\\
\Delta\mu_{\alpha* i}\\ \Delta\mu_{\delta i} \end{array} \right]\, ,
\end{equation}
thus consists of corrections to be added to the \textit{Gaia} parameters. 
With $m$ sources, the total number of parameters to estimate is $6+5m$. 
The estimation is done using a weighted least-squares 
algorithm, using  as ``observations'' the \textit{Gaia} data, hereafter denoted $\vec{g}_i$, and the 
VLBI measurements, denoted $\vec{f}_i$. We now proceed to detail how these observations 
depend on the unknowns.

The general model of the \textit{Gaia} data is
\begin{equation}\label{est01}
\vec{g}_i = \vec{G}_i(\vec{x},\,\vec{y}_i) + \vec{\gamma}_i\, ,
\end{equation}
where $\vec{G}_i$ is a function mapping the model parameters to the expected \textit{Gaia} 
data, according to the model, and $\vec{\gamma}_i$ is the noise. In the standard
five-parameter model, $\vec{g}_i$ and $\vec{\gamma}_i$ are $5\times 1$ column matrices. 
It is assumed that the \textit{Gaia} data are unbiased, except for the frame rotation, and that 
the uncertainties are correctly represented in the \textit{Gaia} catalogue, so that
\begin{equation}\label{eq:gamma}
\text{E}\left[\vec{\gamma}_i\right] = \vec{0} \quad \text{and}\quad
\text{E}\left[\vec{\gamma}_i\vec{\gamma}_i'\right] = \vec{C}_i\, , 
\end{equation}
where $\vec{C}_i$ is the $5\times 5$ covariance matrix of the \textit{Gaia} parameters for source 
$i$. Expressing both $\vec{y}_i$ and $\vec{g}_i$ differentially with respect to the \textit{Gaia} 
values, we have $\vec{g}_i=\vec{0}$ and the linearised version of Eq.~(\ref{est01}) becomes
\begin{equation}\label{est03}
\vec{0} =  \vec{y}_i - \vec{K}_i\vec{x} + \vec{\gamma}_i\, ,
\end{equation}
where
\begin{equation}\label{est02}
\vec{K}_i = \begin{bmatrix} 
\text{c}\alpha_i\text{s}\delta_i & \text{s}\alpha_i\text{s}\delta_i & -\text{c}\delta_i & 0 & 0 & 0 \\
-\text{s}\alpha_i & \text{c}\alpha_i & 0 & 0 & 0 & 0 \\
0 & 0 & 0 & 0 & 0 & 0 \\
0 & 0 & 0 & \text{c}\alpha_i\text{s}\delta_i & \text{s}\alpha_i\text{s}\delta_i & -\text{c}\delta_i\\
0 & 0 & 0 & -\text{s}\alpha_i & \text{c}\alpha_i & 0\\
\end{bmatrix}\, 
\end{equation}
is the matrix containing the trigonometric factors from Eqs.~(\ref{e06})--(\ref{e02}). The third row
of the matrix is zero because the parallax is unaffected by the frame rotation.

Although the linearised form of Eq.~(\ref{est03}), with $\dim(\vec{y}_i)=5$, is used for the rest of this 
paper, the more general expression, Eq.~(\ref{est01}), should be retained for future reference, 
when the \textit{Gaia} observations may provide additional parameters (Sect.~\ref{sec:out}).

The description of the VLBI data for source $i$ is similarly written in the general form
\begin{equation}\label{est04}
\vec{f}_i = \vec{F}_i(\vec{y}_i) + \vec{\nu}_i\, ,
\end{equation}
where $\vec{F}_i$ is a function mapping the source parameters to the expected VLBI data. 
The rotation parameters $\vec{x}$ do not enter here because the VLBI measurements are assumed 
to be in the ICRF frame. The VLBI measurement errors are represented by the column matrix 
$\vec{\nu}_i$, with
\begin{equation}\label{eq:nu}
\text{E}\left[\vec{\nu}_i\right] = \vec{0} \quad \text{and}\quad
\text{E}\left[\vec{\nu}_i\vec{\nu}_i'\right] = \vec{V}_i\, 
\end{equation}
and known covariance matrix $\vec{V}_i$.
The dimension of $\vec{f}_i$ is $n_i\times 1$, where $n_i$ depends on the number of 
VLBI measurements and their state of reduction. If the measurements have been reduced 
to a set of five astrometric parameters, similar to the \textit{Gaia} data but referring to some 
epoch $t_i$ chosen specifically for these observations, we have $n_i=5$. The VLBI data 
could also consist of a single measurement of the topocentric position at epoch $t_i$, however, in 
which case $n_i=2$; or of a sequence of topocentric positions at different epochs. 
In either case, $\vec{F}_i(\vec{y}_i)$ involves a propagation of the source parameters from 
the reference epoch $T$ to the specific epoch(s) of the VLBI data $t_i$. For the standard 
five-parameter astrometric model this propagation should be done as described in 
Appendix~\ref{sec:stand}.

Recalling that $\vec{y}_i=\vec{0}$ represents the source parameters according to \textit{Gaia}, 
we see that $\vec{\Delta f}_i=\vec{f}_i-\vec{F}_i(\vec{0})$ contains the differences between 
the actually observed VLBI data $\vec{f}_i$ and the values $\vec{F}_i(\vec{0})$ computed by 
propagating the \textit{Gaia} parameters to the VLBI epoch. To first order in $\vec{y}_i$, the 
linearised version of Eq.~(\ref{est04}) is therefore
\begin{equation}\label{est05}
\vec{\Delta f}_i = \vec{M}_i \vec{y}_i + \vec{\nu}_i\, ,
\end{equation}
where $\vec{M}_i=\partial\vec{F}_i/\partial\vec{y}_i'$ is the Jacobian matrix evaluated at 
$\vec{y}_i=\vec{0}$. If $\vec{f}_i$ consists of the standard $n_i=5$ astrometric parameters, 
taken in the same order as in Eq.~(\ref{eq:yi}) but referring to epoch $t_i$, then the Jacobian is
approximately given by
\begin{equation}\label{eq:M}
\vec{M}_i\simeq\begin{bmatrix} 1 & 0 & 0 & t_i-T & 0\\ 0 & 1 & 0 & 0 & t_i-T \\
0 & 0 & 1 & 0 & 0 \\ 0 & 0 & 0 & 1 & 0 \\ 0 & 0 & 0 & 0 & 1\end{bmatrix}\, .
\end{equation}
This expression is accurate to  first order in the total proper motion over the time interval 
$t_i-T$, which in some cases could amount to many arcseconds. Because it is then not obvious 
that Eq.~(\ref{eq:M}) is a sufficiently good approximation, it is advisable to evaluate $\vec{M}_i$
by numerical differentiation of the propagation formulae. 

The generalised least-squares estimate is obtained by minimising the loss function
\begin{multline}\label{est06}
Q(\vec{x},\{\vec{y}_i\}_S) = \sum_{i\in S}\, \Bigl[
\left(\vec{y}_i-\vec{K}_i\vec{x}\right)'\vec{C}_i^{-1}\left(\vec{y}_i-\vec{K}_i\vec{x}\right)\\
+\left(\vec{\Delta f}_i-\vec{M}_i\vec{y}_i\right)'\vec{V}_i^{-1}\left(\vec{\Delta f}_i-\vec{M}_i\vec{y}_i\right) 
\Bigr]\, .
\end{multline}
On the assumption of Gaussian errors, the likelihood function is proportional to $\exp(-Q/2)$,
and minimising $Q$ is then equivalent to a maximum-likelihood estimation. 

Setting the partial derivative of $Q$ with respect to each model parameter equal to zero gives the 
symmetric system of $6+5m$ linear equations, known as the normal equations,
\begin{align}
\left(\sum_{i\in S} \vec{K}_i'\vec{C}_i^{-1}\vec{K}_i\right)\vec{x} 
- \sum_{i\in S}\vec{K}_i'\vec{C}_i^{-1}\vec{y}_i &= \vec{0}\, , \label{est07a}\\
-\vec{C}_i^{-1}\vec{K}_i\vec{x}+\left(\vec{C}_i^{-1}+\vec{M}_i'\vec{V}_i^{-1}\vec{M}_i\right)\vec{y}_i
&= \vec{M}_i'\vec{V}_i^{-1}\vec{\Delta f}_i\, , \quad i\in S\, .\label{est07b}
\end{align}
These can be solved by standard numerical methods, and the inverse of the normal 
matrix provides an estimate of the covariance of the model parameters. 

Computationally, the solution of Eqs.~(\ref{est07a})--(\ref{est07b}) is unproblematic 
as it involves only a moderate 
number of unknowns. In terms of numerical accuracy, it is advantageous to compute the 
least-squares solution using orthogonal transformations \citep[e.g.][]{book:bjork-1996}
after transforming the observation equations, Eqs.~(\ref{est03}) and (\ref{est05}), 
to an equivalent set of uncorrelated unit-weight equations. 
Details of this procedure are not given here.

While the least-squares problem is thus solved, there is some additional insight to be
gained by further manipulation of Eqs.~(\ref{est06})--(\ref{est07b}). Using Eq.~(\ref{est07b}), it is 
possible to write each $\vec{y}_i$ in terms of $\vec{x}$; inserting these into Eq.~(\ref{est07a}) 
yields a reduced system of normal equations  involving only the common parameters,%
\footnote{This procedure, known as Helmert blocking after the German geodesist F.~R.~Helmert,
who described the method in 1880 \citep{wolf1978}, is frequently applied to large-scale
least-squares problems in various branches of science, including astrometry
\citep[e.g.][]{1974MNRAS.167..169D}.} 
\begin{equation}\label{est08}
\left(\sum_{i\in S} \vec{N}_i\right) \vec{x} = \sum_{i\in S} \vec{b}_i \, ,
\end{equation}
where
\begin{align}
\vec{N}_i  &= \vec{K}_i'\vec{M}_i'\vec{D}_i^{-1}\vec{M}_i\vec{K}_i \, ,
\label{est09}\\
\vec{b}_i &= \vec{K}_i'\vec{M}_i'\vec{D}_i^{-1}\vec{\Delta f}_i \, ,
\label{est10}
\end{align}
and
\begin{equation}\label{est14}
\vec{D}_i = \vec{V}_i+\vec{M}_i\vec{C}_i\vec{M}_i'\, .
\end{equation}
Solving Eq.~(\ref{est08}) yields $\vec{\hat{x}}$, and then $\vec{\hat{y}}_i$ from the $m$ equations
(\ref{est07b}). Clearly, this solution is mathematically the same as obtained directly from Eqs. 
(\ref{est07a})--(\ref{est07b}). 
It is more remarkable that the covariance of $\vec{\hat{x}}$ 
(the upper left $6\times 6$ submatrix of the inverse of the full normal matrix)
is obtained from the reduced system as $\left(\sum_{i\in S} \vec{N}_i\right)^{-1}$. 

By a similar process of eliminating the unknowns $\vec{y}_i$, the loss function Eq.~(\ref{est06}) 
can be written in terms of $\vec{x}$ as 
\begin{equation}\label{est11}
Q(\vec{x}) = \sum_{i\in S}\, Q_i(\vec{x}) \, ,
\end{equation}
with
\begin{equation}\label{est13}
Q_i(\vec{x})  = \left(\vec{\Delta f}_i-\vec{M}_i\vec{K}_i\vec{x}\right)'\vec{D}_i^{-1}
\left(\vec{\Delta f}_i-\vec{M}_i\vec{K}_i\vec{x}\right) \, .
\end{equation}
The interpretation of the above equations is straightforward. $\vec{\Delta f}_i-\vec{M}_i\vec{K}_i\vec{x}$ 
is the residual of the VLBI data with respect to the values predicted from the \textit{Gaia} data, after
correcting for the rotation parameters and propagating to the VLBI epoch. $\vec{D}_i$ in (\ref{est14})
is the covariance of $\vec{\Delta f}_i$, including the contributions from the uncertainties of both the VLBI 
and propagated \textit{Gaia} data. Equation (\ref{est13}) shows that $\vec{\hat{x}}$ minimises the
sum of the squares of the VLBI residuals after normalisation by the combined uncertainties.
For a given solution $\vec{\hat{x,}}$ we may take the quantity 
\begin{equation}\label{est13a}
Q_i/n_i = Q_i(\vec{\hat{x}})/\!\dim(\vec{f}_i)
\end{equation}
as a measure of the discrepancy for source $i$, where $n_i=\dim(\vec{f}_i)$ is the number of 
VLBI data points included for the source. The normalisation by $n_i$ is essential in order to avoid 
penalising sources with many VLBI data points. $Q_i/n_i$ can be interpreted as the reduced 
chi-square of the source, and should ideally be around unity if the astrometric model fits the 
source and the uncertainties are correctly estimated. $Q_i/n_i$ is hereafter referred to as the 
``discrepancy measure'' of the source relative to a given solution.

Equations (\ref{est08})--(\ref{est13}) have some practical advantages over the use of orthogonal 
transformations to solve the least-squares problem. For the identification of outliers (Sect.~\ref{sec:out}),
computing the solution and other statistics for a very large number of different
subsets of $S$ may be required. This can be done most efficiently by pre-computing $\vec{D}_i$, 
$\vec{N}_i$, $\vec{b}_i,$ and other quantities that do not depend on the solution. 

The matrices $\vec{N}_i$ are, furthermore, useful for quantifying the amount of (Fisher) information 
on $\vec{x}$ contributed by each source. A source without VLBI data would formally have infinite 
$\vec{V}_i$ and hence $\vec{N}_i=\vec{0}$. A source with only proper motion data from VLBI
will not contribute to the estimation of $\vec{\varepsilon}$ and will consequently have zeros in the 
first three rows and columns of $\vec{N}_i$. More generally, the amount of information contributed 
by source $i$ to the estimation of $\vec{\varepsilon}(T)$ and $\vec{\omega}$ is quantified by
\begin{equation}\label{est15}
E_i = \underset{\vec{\varepsilon}}{\text{trace}}(\vec{N}_i) \quad\text{and}\quad
\Omega_i = \underset{\vec{\omega}}{\text{trace}}(\vec{N}_i) \, ,
\end{equation} 
respectively, where the trace is limited to the first three diagonal elements of $\vec{N}_i$ for 
$E_i$, and to the last three for $\Omega_i$.

\subsection{Modelling issues and robustness}
\label{sec:out}

In the preceding treatment it was assumed that the sources move through space with uniform velocity 
relative to the Solar System barycentre, allowing both the \textit{Gaia} and VLBI measurements to
be accurately modelled by five astrometric parameters per source, plus a spectroscopically 
determined radial velocity. This model is manifestly incorrect for a number of radio stars known to be 
members of binaries or more complex systems, for which the VLBI observations have determined 
non-linear motions \citep[e.g.\ the T~Tau system;][]{2007ApJ...671..546L} or even complete orbits
\citep[e.g.\ the Algol and UX~Arietis systems;][]{2011ApJ...737..104P}. A second assumption is that
the centre of radio emission coincides with that of the optical emission, which is also not true for many
objects with extended atmospheres, discs, jets, and other structures in the radio and/or optical images.
The astrometric biases produced by these various effects range over many decades, from the
undetectable to tens of milliarcseconds. As a result, the simple modelling described above will provide excellent 
fits for some sources and large residuals for others, with a continuum of intermediate cases. 

The general method of estimation in Sect.~\ref{sec:est} does permit the application of more 
sophisticated source models. Depending on the physical nature of a radio star, it may be possible 
to improve the modelling, and ultimately the accuracy of $\vec{\hat{x}}$, by introducing a small 
number of additional unknowns, thus extending the array $\vec{y}_i$ in Eq.~(\ref{eq:yi}). For example,
in an interacting binary it may be possible to model the offsets of the optical and radio emissions 
from the barycentre of the system in terms of a few geometrical elements if the main characteristics 
of the binary, including its period, are known from spectroscopy. Another example is the non-linear 
motion of a radio star perturbed by a distant companion. In this case, the comparison of radio and
optical observations, taken at different epochs, is meaningful only if the non-linearity is taken
into account through the addition of a few acceleration terms or possibly a complete set of 
orbital elements. In either case, the modelling requires an augmented parameter array $\vec{y}_i$ 
and more elaborate expressions for the functions $\vec{G}_i$ and $\vec{F}_i$ in Eqs.~(\ref{est01}) and (\ref{est04}) than discussed in Sect.~\ref{sec:est}. However, the details of this are beyond the 
scope of this paper, where the standard model of Appendix~\ref{sec:stand} is used throughout.

To cope with unmodelled effects, whether they are radio--optical offsets, non-linear motions, or
deviations from more elaborate models, it is imperative that the estimation procedure is robust, 
that is,\ that the result is not overly sensitive to the relatively few cases with large perturbations. 
The least-squares estimation of Sect.~\ref{sec:est} is inherently non-robust and needs 
to be modified or complemented with other techniques to provide the required robustness. 
The strategy adopted in this paper is to identify the most problematic sources and exclude them 
from the solution. Consistent with the treatment in Sect.~\ref{sec:est}, each source with all its
data is regarded as an independent entity, to be either included or rejected. Although it could 
happen that the model fits the data very well in one coordinate (say, $\alpha$), but not in the 
other ($\delta$), a reasonable standpoint is that such a source is better left out 
entirely. Discrepant sources can be identified by means of statistics such as Eq.~(\ref{est13a}), and
by comparing solutions for different subsets of $S$. Details of the procedure are explained in 
Sect.~\ref{sec:result} as it is applied to actual data.

\begin{table*}
\caption{Astrometric parameters determined by VLBI for the radio sources included in the analysis.}
\centering\tiny
\setlength{\tabcolsep}{2pt}
\begin{tabular}{llrcrrrrrr}
\hline\hline\noalign{\medskip}
Name 
& Type 
& \multicolumn{1}{c}{$v_r$} 
& \multicolumn{1}{c}{Epoch $t$} 
& \multicolumn{1}{c}{$\alpha(t)$} 
& \multicolumn{1}{c}{$\delta(t)$} 
& \multicolumn{1}{c}{$\varpi(t)$} 
& \multicolumn{1}{c}{$\mu_{\alpha*}(t)$} 
& \multicolumn{1}{c}{$\mu_{\delta}(t)$} 
& \multicolumn{1}{c}{Ref.} 
\\
& 
& \multicolumn{1}{c}{(km~s$^{-1}$)} 
& \multicolumn{1}{c}{(Julian year)} 
& \multicolumn{1}{c}{($\text{deg}\pm\text{mas}$)} 
& \multicolumn{1}{c}{($\text{deg}\pm\text{mas}$)} 
& \multicolumn{1}{c}{(mas)} 
& \multicolumn{1}{c}{(mas~yr$^{-1}$)} 
& \multicolumn{1}{c}{(mas~yr$^{-1}$)} 
& \multicolumn{1}{c}{} 
\\
\noalign{\smallskip}\hline\noalign{\medskip}
{}SY Scl & Mira & $ 24.0$ & 2006.8282 & $  1.901028361$~$\pm15.10$ & $-25.494452258$~$\pm7.700$ & $ 0.750$~$\pm0.030$ & $   5.570$~$\pm0.040$ & $  -7.320$~$\pm0.120$ & 1\\
{}S Per & RedSG & $-39.7$ & 2000.8884 & $ 35.715460833$~$\pm7.818$ & $ 58.586512222$~$\pm8.000$ & $ 0.413$~$\pm0.017$ & $  -0.490$~$\pm0.230$ & $  -1.190$~$\pm0.200$ & 2\\
{}LS I +61 303 & HXB & $-41.4$ & 1992.0000 & $ 40.131935007$~$\pm0.291$ & $ 61.229332410$~$\pm0.573$ & $ 0.260$~$\pm0.610$ & $   0.967$~$\pm0.260$ & $  -1.210$~$\pm0.320$ & 3\\
{}UX Ari & RSCVn & $ 50.7$ & 1999.9863 & $ 51.647432750$~$\pm1.200$ & $ 28.715076528$~$\pm0.800$ & $19.900$~$\pm0.500$ & $  44.960$~$\pm0.130$ & $-102.330$~$\pm0.090$ & 4\\
$\cdots$ &$\cdots$ &$\cdots$ & 1991.2471 & $ 51.647339076$~$\pm0.331$ & $ 28.715340363$~$\pm0.377$ & $19.370$~$\pm0.390$ & $  41.617$~$\pm0.186$ & $-104.010$~$\pm0.200$ & 3\\
{}HD 22468 & RSCVn & $-15.3$ & 1992.0000 & $ 54.197113376$~$\pm0.406$ & $  0.588121131$~$\pm0.401$ & $33.880$~$\pm0.470$ & $ -31.588$~$\pm0.330$ & $-161.690$~$\pm0.310$ & 3\\
{}V1271 Tau & RSCVn & $  5.5$ & 2013.0000 & $ 55.951464140$~$\pm0.110$ & $ 25.004223050$~$\pm0.171$ & $ 7.418$~$\pm0.025$ & $  19.860$~$\pm0.050$ & $ -45.410$~$\pm0.160$ & 5\\
{}V811 Tau & BYDra & $ 11.3$ & 2013.0000 & $ 56.338353995$~$\pm0.118$ & $ 23.727308022$~$\pm0.198$ & $ 7.223$~$\pm0.057$ & $  19.470$~$\pm0.110$ & $ -44.390$~$\pm0.270$ & 5\\
{}HD 283447 & TTau & $ 16.0$ & 2006.9700 & $ 63.553843529$~$\pm0.132$ & $ 28.203380745$~$\pm0.097$ & $ 7.700$~$\pm0.190$ & $  17.092$~$\pm0.077$ & $ -24.030$~$\pm0.053$ & 6\\
$\cdots$ &$\cdots$ &$\cdots$ & 1993.8809 & $ 63.553831265$~$\pm0.312$ & $ 28.203455247$~$\pm0.470$ & $ 6.740$~$\pm0.250$ & $   0.423$~$\pm0.291$ & $ -23.250$~$\pm0.280$ & 3\\
{}V410 Tau & TTau & $ 19.9$ & 2015.7600 & $ 64.629661292$~$\pm0.021$ & $ 28.454377914$~$\pm0.034$ & $ 7.751$~$\pm0.027$ & $   8.703$~$\pm0.017$ & $ -24.985$~$\pm0.020$ & 7\\
{}V1023 Tau & TTau & $ 12.7$ & 2005.3525 & $ 64.695968392$~$\pm0.013$ & $ 28.335383111$~$\pm0.200$ & $ 7.530$~$\pm0.030$ & $   4.300$~$\pm0.050$ & $ -28.900$~$\pm0.300$ & 8\\
{}HD 283572 & TTau & $ 14.2$ & 2005.3525 & $ 65.495216792$~$\pm0.264$ & $ 28.301769800$~$\pm0.050$ & $ 7.780$~$\pm0.040$ & $   8.880$~$\pm0.060$ & $ -26.600$~$\pm0.100$ & 8\\
$\cdots$ &$\cdots$ &$\cdots$ & 2005.3600 & $ 65.495216803$~$\pm0.032$ & $ 28.301769751$~$\pm0.035$ & $ 7.841$~$\pm0.057$ & $   9.023$~$\pm0.061$ & $ -26.445$~$\pm0.077$ & 7\\
{}T Tau & TTau & $ 23.9$ & 2004.6231 & $ 65.497604438$~$\pm0.028$ & $ 19.534921017$~$\pm0.400$ & $ 6.820$~$\pm0.030$ & $   4.020$~$\pm0.030$ & $  -1.180$~$\pm0.050$ & 9\\
{}HD 283641 & TTau & $ 16.2$ & 2015.9500 & $ 66.204409070$~$\pm0.065$ & $ 26.719480101$~$\pm0.085$ & $ 6.285$~$\pm0.070$ & $  10.913$~$\pm0.037$ & $ -16.772$~$\pm0.044$ & 7\\
{}V1110 Tau & RSCVn & $   $ & 2015.5700 & $ 68.663414796$~$\pm0.109$ & $ 25.016900201$~$\pm0.116$ & $11.881$~$\pm0.149$ & $ -52.705$~$\pm0.062$ & $ -11.321$~$\pm0.066$ & 7\\
{}HD 282630 & TTau & $ 13.6$ & 2016.5800 & $ 73.904063653$~$\pm0.067$ & $ 30.298527621$~$\pm0.066$ & $ 7.061$~$\pm0.125$ & $   3.897$~$\pm0.113$ & $ -24.210$~$\pm0.132$ & 7\\
{}T Lep & Mira & $ -4.0$ & 2005.6427 & $ 76.211833879$~$\pm30.18$ & $-21.904579275$~$\pm13.15$ & $ 3.060$~$\pm0.040$ & $  14.600$~$\pm0.500$ & $ -35.430$~$\pm0.790$ & 10\\
{}V1961 Ori & Orion\_V* & $ 30.1$ & 2014.8600 & $ 83.613911148$~$\pm0.027$ & $ -5.406194168$~$\pm0.028$ & $ 2.533$~$\pm0.027$ & $  -7.220$~$\pm0.060$ & $  -0.990$~$\pm0.080$ & 11\\
{}Brun 334 & PMS & $ 21.3$ & 2015.1800 & $ 83.665666567$~$\pm0.023$ & $ -5.407112221$~$\pm0.048$ & $ 2.591$~$\pm0.046$ & $  -4.010$~$\pm0.080$ & $  -1.170$~$\pm0.070$ & 11\\
{}V1321 Ori & Orion\_V* & $ 17.6$ & 2015.1800 & $ 83.767921447$~$\pm0.075$ & $ -5.136841216$~$\pm0.182$ & $ 2.509$~$\pm0.044$ & $   0.060$~$\pm0.200$ & $   6.950$~$\pm0.160$ & 11\\
{}MT Ori & Orion\_V* & $   $ & 2015.1800 & $ 83.824801264$~$\pm0.030$ & $ -5.379288056$~$\pm0.062$ & $ 2.646$~$\pm0.041$ & $   3.820$~$\pm0.100$ & $   1.600$~$\pm0.170$ & 11\\
{}V1046 Ori & SB & $ 29.5$ & 2015.2000 & $ 83.841117796$~$\pm0.054$ & $ -4.494167133$~$\pm0.121$ & $ 2.643$~$\pm0.075$ & $   1.880$~$\pm0.090$ & $   1.200$~$\pm0.140$ & 11\\
{}HD 37150 & Star & $ 10.8$ & 2015.1900 & $ 84.062621055$~$\pm0.045$ & $ -5.647921674$~$\pm0.096$ & $ 2.536$~$\pm0.046$ & $   1.320$~$\pm0.050$ & $  -0.560$~$\pm0.120$ & 11\\
{}TYC 5346-538-1 & Star & $   $ & 2015.2000 & $ 85.640319910$~$\pm0.058$ & $ -8.120884301$~$\pm0.140$ & $ 2.348$~$\pm0.069$ & $   0.680$~$\pm0.090$ & $  -0.510$~$\pm0.250$ & 11\\
{}HD 290862 & Star & $   $ & 2015.2100 & $ 86.680772978$~$\pm0.491$ & $  0.076677687$~$\pm0.643$ & $ 2.197$~$\pm0.545$ & $   0.350$~$\pm0.270$ & $   0.830$~$\pm0.830$ & 11\\
{}[SSC75] M 78 11 & Star & $   $ & 2015.2100 & $ 86.688907482$~$\pm0.043$ & $  0.044527387$~$\pm0.085$ & $ 2.547$~$\pm0.034$ & $   0.010$~$\pm0.100$ & $  -0.490$~$\pm0.080$ & 11\\
{}VY CMa & RedSG & $ 60.8$ & 2006.5300 & $110.743024583$~$\pm10.00$ & $-25.767517500$~$\pm10.00$ & $ 0.830$~$\pm0.080$ & $  -2.210$~$\pm0.060$ & $   2.290$~$\pm0.300$ & 12\\
{}S Crt & LPV & $ 32.0$ & 2005.7988 & $178.187373874$~$\pm1.800$ & $ -7.596690803$~$\pm13.30$ & $ 2.330$~$\pm0.130$ & $  -3.170$~$\pm0.220$ & $  -5.410$~$\pm0.220$ & 13\\
{}BH CVn & RSCVn & $  6.4$ & 1993.1088 & $203.698997828$~$\pm0.373$ & $ 37.182433334$~$\pm0.455$ & $22.210$~$\pm0.450$ & $  85.496$~$\pm0.131$ & $  -9.220$~$\pm0.160$ & 3\\
{}S CrB & Mira & $ -5.1$ & 2000.0000 & $230.349817010$~\phantom{$\pm0.000$} & $ 31.367381400$~\phantom{$\pm0.000$} & $ 2.360$~$\pm0.230$ & $  -9.060$~$\pm0.230$ & $ -12.520$~$\pm0.290$ & 14\\
{}$\sigma^2$ CrB & HPM & $-14.8$ & 1990.0014 & $243.671114605$~$\pm0.104$ & $ 33.858853887$~$\pm0.124$ & $43.930$~$\pm0.100$ & $-267.048$~$\pm0.037$ & $ -86.660$~$\pm0.050$ & 3\\
{}U Her & Mira & $-26.1$ & 2000.0000 & $246.447798640$~\phantom{$\pm0.000$} & $ 18.892459900$~\phantom{$\pm0.000$} & $ 3.740$~$\pm0.610$ & $ -14.980$~$\pm0.290$ & $  -9.230$~$\pm0.320$ & 14\\
{}Haro 1-6 & TTau & $   $ & 2007.9900 & $246.512566654$~$\pm0.399$ & $-24.393446456$~$\pm0.356$ & $ 7.385$~$\pm0.234$ & $ -19.630$~$\pm0.190$ & $ -26.920$~$\pm0.130$ & 15\\
{}DoAr 51 & TTau & $   $ & 2014.7500 & $248.049128879$~$\pm0.337$ & $-24.672768034$~$\pm0.336$ & $ 6.983$~$\pm0.050$ & $  -4.800$~$\pm0.080$ & $ -23.110$~$\pm0.110$ & 15\\
{}W 40 IRS 5 & Star & $   $ & 2014.9500 & $277.811761649$~$\pm0.065$ & $ -2.063932732$~$\pm0.119$ & $ 2.302$~$\pm0.063$ & $   0.186$~$\pm0.053$ & $  -6.726$~$\pm0.121$ & 16\\
{}RR Aql & Mira & $ 11.0$ & 2000.0000 & $299.400248870$~\phantom{$\pm0.000$} & $ -1.886482960$~\phantom{$\pm0.000$} & $ 1.580$~$\pm0.400$ & $ -25.110$~$\pm0.740$ & $ -49.820$~$\pm0.540$ & 14\\
{}Cyg X-1 & HXB & $ -2.7$ & 2009.5700 & $299.590303145$~$\pm0.500$ & $ 35.201590284$~$\pm0.500$ & $ 0.547$~$\pm0.041$ & $  -3.700$~$\pm0.080$ & $  -6.420$~$\pm0.140$ & 17\\
$\cdots$ &$\cdots$ &$\cdots$ & 1991.2498 & $299.590326964$~$\pm0.308$ & $ 35.201622230$~$\pm0.368$ & $ 0.730$~$\pm0.300$ & $  -3.787$~$\pm0.172$ & $  -6.250$~$\pm0.210$ & 3\\
{}HD 199178 & RVS & $-26.6$ & 1993.7933 & $313.473486712$~$\pm0.332$ & $ 44.386412565$~$\pm0.397$ & $ 8.590$~$\pm0.330$ & $  26.595$~$\pm0.407$ & $  -1.240$~$\pm0.430$ & 3\\
{}SS Cyg & DwarfNova & $-62.0$ & 2011.5661 & $325.678846337$~$\pm0.065$ & $ 43.586181392$~$\pm0.070$ & $ 8.800$~$\pm0.120$ & $ 112.420$~$\pm0.070$ & $  33.380$~$\pm0.070$ & 18\\
{}AR Lac & RSCVn & $-33.8$ & 1992.4353 & $332.170232910$~$\pm0.274$ & $ 45.742153188$~$\pm0.361$ & $23.970$~$\pm0.370$ & $ -52.080$~$\pm0.126$ & $  47.030$~$\pm0.190$ & 3\\
{}IM Peg & RSCVn & $-14.4$ & 2005.0869 & $343.259410883$~$\pm0.400$ & $ 16.841155569$~$\pm0.390$ & $10.370$~$\pm0.074$ & $ -20.833$~$\pm0.090$ & $ -27.267$~$\pm0.095$ & 19\\
$\cdots$ &$\cdots$ &$\cdots$ & 1992.9172 & $343.259484359$~$\pm0.360$ & $ 16.841247861$~$\pm0.392$ & $10.280$~$\pm0.620$ & $ -20.587$~$\pm0.459$ & $ -27.530$~$\pm0.400$ & 3\\
{}PZ Cas & RedSG & $-51.4$ & 2006.2998 & $356.013673333$~$\pm2.836$ & $ 61.789496389$~$\pm3.000$ & $ 0.356$~$\pm0.026$ & $  -3.700$~$\pm0.200$ & $  -2.000$~$\pm0.300$ & 20\\
\noalign{\smallskip}\hline
\end{tabular}\label{table1}
\tablefoot{Name, type, and radial velocity ($v_r$) are taken from SIMBAD, 
except for the names Cyg~X-1 (for HD~226868) and $\sigma^2$~CrB (for sig~CrB~A).
An ellipsis ($\cdots$) means the same data as in the line above. 
Positions $\alpha(t)$, $\delta(t)$ are barycentric and refer to the epochs in the fourth
column. Uncertainties (in mas and mas~yr$^{-1}$) are given after the $\pm$ sign; for $\alpha$
they are $\sigma_{\alpha*}=\sigma_\alpha\cos\delta$. 
Positional data without uncertainties were not used in the solutions. 
(1)~\citet{2011PASJ...63...63N};
(2)~\citet{2010ApJ...721..267A};
(3)~\citet{1999A&A...344.1014L};
(4)~\citet{2011ApJ...737..104P};
(5)~\citet{2014Sci...345.1029M};
(6)~\citet{2012ApJ...747...18T};
(7)~\citet{2018ApJ...859...33G};
(8)~\citet{2007ApJ...671.1813T};
(9)~\citet{2007ApJ...671..546L};
(10)~\citet{2014PASJ...66..101N};
(11)~\citet{2017ApJ...834..142K};
(12)~\citet{2012ApJ...744...23Z};
(13)~\citet{2008PASJ...60.1013N};
(14)~\citet{2007A&A...472..547V};
(15)~\citet{2017ApJ...834..141O};
(16)~\citet{2017ApJ...834..143O};
(17)~\citet{2011ApJ...742...83R};
(18)~\citet{2013Sci...340..950M};
(19)~\citet{2015CQGra..32v4021B};
(20)~\citet{2013ApJ...774..107K}.
}
\end{table*}

\begin{table*}
\caption{\textit{Gaia} DR2 matches and solution statistics for the radio sources in Table~1.}
\centering
\footnotesize
\begin{tabular}{lrrrrrrrc}
\hline\hline\noalign{\medskip}
Name 
& \multicolumn{1}{c}{\textit{Gaia} DR2 identifier} 
& \multicolumn{1}{c}{$G$} 
& \multicolumn{1}{c}{RUWE} 
& \multicolumn{1}{c}{$E_i$} 
& \multicolumn{1}{c}{$\Omega_i$} 
& \multicolumn{1}{r}{$n_i$} 
& \multicolumn{1}{r}{$Q_i/n_i$} 
& \multicolumn{1}{c}{Accepted} 
\\
& 
& \multicolumn{1}{c}{(mag)} 
& 
& \multicolumn{1}{c}{(mas$^{-2}$)} 
& \multicolumn{1}{c}{(mas$^{-2}$~yr$^2$)} 
& &&\multicolumn{1}{c}{(0/1)} 
\\
\noalign{\smallskip}\hline\noalign{\medskip}
{}SY Scl & 2335529621301024128 & 10.09 &  1.28 &    0.0~~ &   20.2~~~ & 5 &      2.556 & 1\\
{}S Per & 459101393719884800 &  7.80 &  1.27 &    0.0~~ &   16.6~~~ & 5 &      4.246 & 1\\
{}LS I +61 303 & 465645515129855872 & 10.39 &  0.92 &    1.4~~ &  755.1~~~ & 5 &      9.823 & 1\\
{}UX Ari & 118986060277836160 &  6.33 &  6.33 &    0.7~~ &   12.1~~~ & 10 &    766.906 & 0\\
{}HD 22468 & 3263936692671872384 &  5.60 &  1.01 &    0.3~~ &  143.7~~~ & 5 &     11.515 & 1\\
{}V1271 Tau & 69876712724339456 & 11.44 &  1.15 &   54.6~~ &  197.4~~~ & 5 &      0.467 & 1\\
{}V811 Tau & 65190048708380160 & 12.09 &  1.08 &   42.1~~ &  240.4~~~ & 5 &      4.684 & 1\\
{}HD 283447 & 163184366130809984 &  9.98 &  3.99 &   21.8~~ &   23.1~~~ & 10 &  19983.091 & 0\\
{}V410 Tau & 164518589131083136 & 10.32 &  0.98 & 1156.3~~ &  339.1~~~ & 5 &      0.915 & 1\\
{}V1023 Tau & 164513538249595136 & 11.65 &  3.27 &    3.9~~ &   27.1~~~ & 5 &   1704.823 & 0\\
{}HD 283572 & 164536250037820160 &  8.80 &  1.02 &   10.2~~ &  273.7~~~ & 10 &      1.273 & 1\\
{}T Tau & 48192969034959232 &  9.63 &  1.69 &   12.0~~ &  165.6~~~ & 5 & 255047.336 & 0\\
{}HD 283641 & 152104381299305600 & 10.80 &  1.62 &  210.6~~ &  110.1~~~ & 5 &      7.636 & 1\\
{}V1110 Tau & 147940663906992512 &  9.96 &  0.90 &  147.6~~ &  203.3~~~ & 5 &     88.639 & 0\\
{}HD 282630 & 156900622818205312 & 10.33 &  1.02 &  255.1~~ &  260.1~~~ & 5 &     73.700 & 0\\
{}T Lep & 2962625495403737600 &  6.44 &  1.39 &    0.0~~ &    5.3~~~ & 5 &     73.666 & 0\\
{}V1961 Ori & 3209424108758593408 & 11.61 &  1.09 &  605.3~~ &  201.2~~~ & 5 &      5.042 & 1\\
{}Brun 334 & 3017270879709003520 & 10.82 &  1.10 &  699.3~~ &  169.8~~~ & 5 &      4.894 & 1\\
{}V1321 Ori & 3209531650444835840 & 10.13 &  0.96 &  176.8~~ &   72.1~~~ & 5 &      2.995 & 1\\
{}MT Ori & 3017364127743299328 & 11.32 &  1.16 &  479.0~~ &   97.3~~~ & 5 &     71.869 & 0\\
{}V1046 Ori & 3209634905754969856 &  6.53 &  2.21 &  100.1~~ &   14.8~~~ & 5 &    127.846 & 0\\
{}HD 37150 & 3017294622299833216 &  6.51 &  0.98 &  294.4~~ &   95.9~~~ & 5 &     62.206 & 0\\
{}TYC 5346-538-1 & 3015742318025842944 & 10.65 &  0.93 &  205.8~~ &   59.6~~~ & 5 &      8.258 & 1\\
{}HD 290862 & 3219148872492984192 & 10.18 &  1.65 &    6.4~~ &   13.6~~~ & 5 &     19.693 & 1\\
{}[SSC75] M 78 11 & 3219148185299243776 & 12.51 &  1.10 &  371.8~~ &  167.7~~~ & 5 &    702.849 & 0\\
{}VY CMa & 5616197646838713728 &  7.17 & 17.20 &    0.0~~ &    0.7~~~ & 5 &     43.772 & 0\\
{}S Crt & 3595101382979440256 &  6.42 &  1.40 &    0.2~~ &   27.5~~~ & 5 & 124262.639 & 0\\
{}BH CVn & 1475118788534734592 &  4.73 &  0.85 &    0.3~~ &   77.4~~~ & 5 &      2.213 & 1\\
{}S CrB & 1277100833181122816 &  7.40 &  1.70 &    0.0~~ &    8.4~~~ & 3 &      5.828 & 1\\
{}$\sigma^2$ CrB & 1328866562170960512 &  5.41 &  1.14 &    1.9~~ &  158.2~~~ & 5 &     65.332 & 0\\
{}U Her & 1200834239913483392 &  7.04 &  1.35 &    0.0~~ &   14.3~~~ & 3 &      5.294 & 1\\
{}Haro 1-6 & 6049142032584969088 & 12.24 &  0.96 &    2.6~~ &   91.7~~~ & 5 &      1.725 & 1\\
{}DoAr 51 & 6047570826172040960 & 12.52 &  6.43 &   10.2~~ &    3.2~~~ & 5 &      7.127 & 1\\
{}W 40 IRS 5 & 4270236599432306560 & 12.13 &  2.98 &   46.5~~ &   15.4~~~ & 5 &  17316.589 & 0\\
{}RR Aql & 4234448531043979520 &  8.18 &  1.55 &    0.0~~ &    4.0~~~ & 3 &     12.770 & 1\\
{}Cyg X-1 & 2059383668236814720 &  8.52 &  0.98 &    6.3~~ &  736.5~~~ & 10 &      1.122 & 1\\
{}HD 199178 & 2162964329341318656 &  7.01 &  0.88 &    1.0~~ &  463.1~~~ & 5 &      0.573 & 1\\
{}SS Cyg & 1972957892448494592 & 11.69 &  1.42 &   32.9~~ &  181.1~~~ & 5 &     11.217 & 1\\
{}AR Lac & 1962909425622345728 &  5.89 &  0.82 &    1.9~~ &  933.1~~~ & 5 &      2.119 & 1\\
{}IM Peg & 2829193299742131328 &  5.66 &  1.43 &    4.0~~ &  115.3~~~ & 10 &      1.233 & 1\\
{}PZ Cas & 2012859963999694848 &  6.64 &  1.06 &    0.2~~ &   40.9~~~ & 5 &     28.604 & 1\\
\noalign{\smallskip}\hline
\end{tabular}\label{table2}
\tablefoot{The quantity $G$ is the mean magnitude in the \textit{Gaia} photometric band; 
RUWE is the re-normalised unit weight error \citep{LL:LL-124} of the astrometric solution 
for the source in \textit{Gaia} DR2. The quantities 
$E_i$ and $\Omega_i$ are the formal weights potentially contributed by the source to the 
estimation of $\vec{\varepsilon}(T)$ and $\vec{\omega}$, computed using Eq.~(29), and
$n_i$ the number of VLBI data items on the source.
The second to last column is the discrepancy measure from Eq.~(28), and the last column
tells whether the source was accepted (1) or not (0) in the baseline solution.}
\end{table*}

\section{Application to \textit{Gaia} DR2 data}
\label{sec:demo}

In this section we use the algorithms in Sect.~\ref{sec:est} to estimate the
rotation parameters of the bright \textit{Gaia} DR2 reference frame, based on VLBI astrometry
of radio stars collected from the literature. The primary goal is to verify, if possible, the
spin detected from comparison with \textsc{Hipparcos} data (Sect.~\ref{sec:intro}), 
but an important secondary goal
is to illustrate the usefulness of positional VLBI data for estimating the spin, compared
with a solution using only proper motions.

\subsection{VLBI data}
\label{sec:vlbi}

Recent technological advances have dramatically expanded the scope for stellar radio
astrophysics \citep{2019PASP..131a6001M}. The amount of accurate VLBI astrometry that could
potentially be used for validating the stellar reference frame of \textit{Gaia} increases rapidly, not least thanks to a number of surveys aiming to study Galactic structure
(e.g.\ BeSSeL, \citealt{2011AN....332..461B}; VERA, \citealt{2013ASPC..476...81H}; GOBELINS, 
\citealt{2018ApJ...865...73O}). A recent comparison of \textit{Gaia} DR2 and VLBI stellar parallaxes 
\citep{2019arXiv190304105X} lists more than 100 targets. 

\begin{figure}[t]
\centerline{
\resizebox{\hsize}{!}{\includegraphics{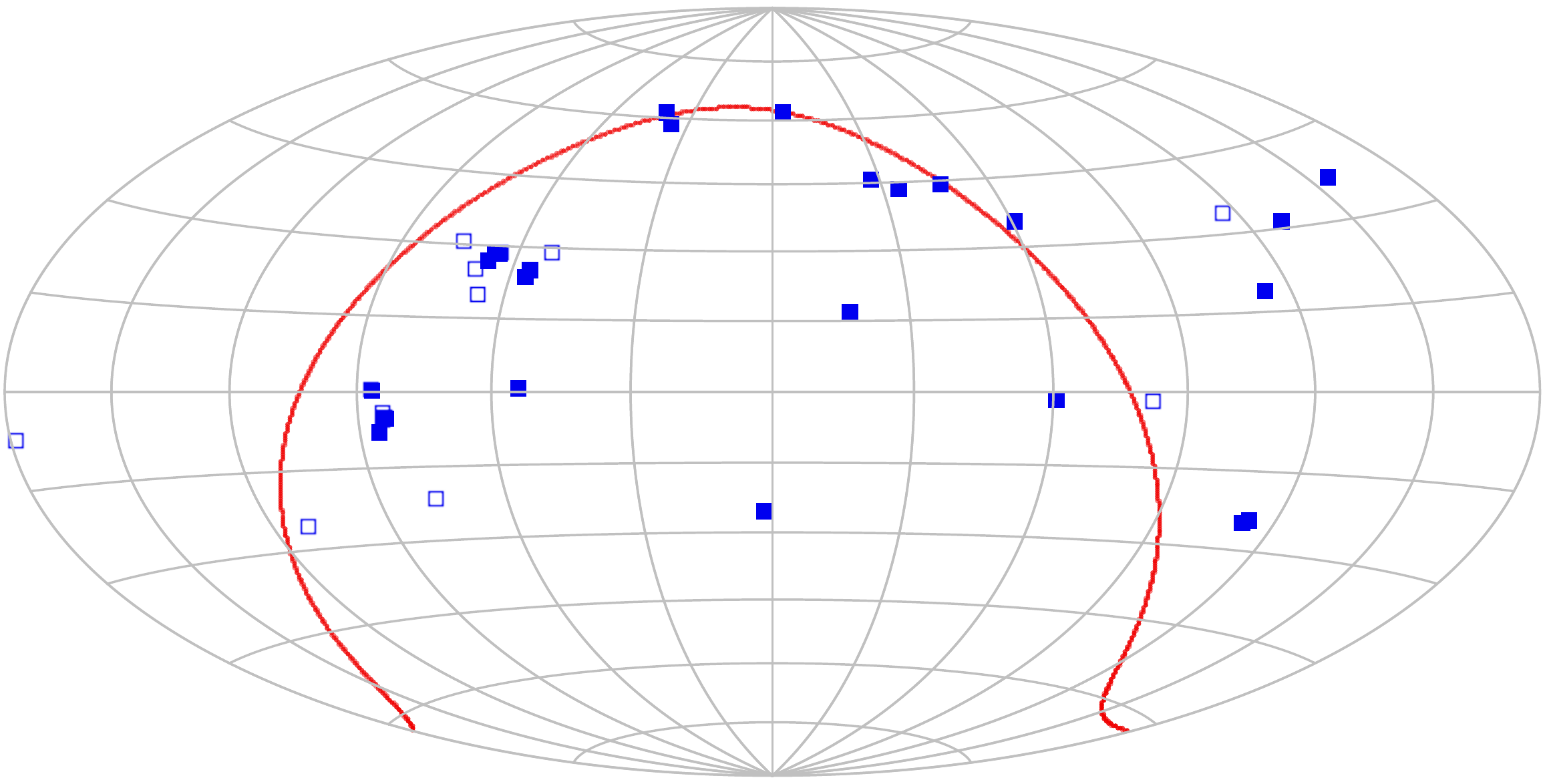}}
}
\caption{Sky distribution of the radio sources in Tables~\ref{table1} and \ref{table2}.
Filled and open symbols mark the objects that were accepted and rejected in the baseline solution.
The map is a Hammer--Aitoff projection in ICRS coordinates with $\alpha=\delta=0$ in the centre, 
$\alpha$ increasing from right to left, and the Galactic equator is plotted in red.
\label{fig02}}
\end{figure}

The 41 objects considered in this study are listed in Table~\ref{table1}, with their corresponding 
\textit{Gaia} DR2 identifiers in Table~\ref{table2}. All the sources are brighter than $G=13.0$ 
and have full (five-parameter) astrometry in \textit{Gaia} DR2. The list includes most radio stars 
from the early programmes, in particular, \citet{1999A&A...344.1014L}, in view of their potentially 
high weight in the estimation of the spin components. These were complemented by a selection 
of more recent data mainly from the GOBELINS survey, which have observation epochs close to 
the \textit{Gaia} DR2 epoch and therefore could contribute usefully to the estimation
of the frame orientation at J2015.5. Most of the objects are young stellar objects, interacting 
binaries, or giants with extended atmospheres. Their celestial distribution is shown in Fig.~\ref{fig02}.

The VLBI data have been collected from some 20 different publications as listed in Table~\ref{table1}. 
In some cases, the authors did not publish the barycentric position at mean epoch obtained in 
their analysis of the observations. In most of these cases, the authors did provide 
results from their individual VLBI sessions, however, including the dates and geocentric positions, from 
which the required information could be reconstructed (cf.\ Sect.~\ref{sec:discPlea}). 
For the sources with reference number 5, 7, 11, 15, and 16 in the last column of Table~\ref{table1},
this is how the listed barycentric positions $\alpha(t)$ and $\delta(t)$ were derived, with
their uncertainties, while the parallaxes and proper motions were taken from the 
cited references. 

For Mira variables and red supergiants, the VLBI observations locate several maser spots 
in the very extended ($\sim\,$100~mas) circumstellar envelopes. When a kinematic model 
is employed for the relative motions of the spots caused by expansion and rotation of the 
envelope, it is often possible to infer the position of the geometrical centre of the star 
\citep[e.g.][]{2012ApJ...744...23Z,2014PASJ...66..101N}. The VLBI positions of these objects 
given in Table~\ref{table1} are not accurate enough to contribute significantly to the determination
of the rotation parameters, but the systemic proper motions may contribute to the spin.  
For reference number 14, the positions in Table~\ref{table1} were taken from SIMBAD and 
are not used in the solutions except to compute the trigonometric factors in Eq.~(\ref{est02}).  

For some older observation series, in particular, \citet[][reference number 3]{1999A&A...344.1014L}, 
the original results have been updated 
using more accurate calibrator (quasar) positions from the ICRF3 catalogue,%
\footnote{\url{https://www.iers.org/IERS/EN/DataProducts/ICRF/ICRF3/icrf3.html}, using the S/X data.
In this process the effect of the Galactic acceleration on the quasar positions was neglected.}  
resulting in some considerable improvements. An example is Cyg~X-1 
(HD~226868), where the positional uncertainties at 1991.25 were reduced from 
$(\sigma_{\alpha*},\,\sigma_\delta)=(1.24,\, 1.75)$~mas, as given in \citet{1999A&A...344.1014L}, 
to the $(0.308,\,0.368)$~mas of Table~\ref{table1}.

In Eq.~(\ref{est05}) the VLBI data are compared with the \textit{Gaia} data propagated to 
the epoch ($t$) of the VLBI data. This was done using the formulae in Appendix~\ref{sec:stand}, 
which requires knowledge of the radial motion of the source in order to take perspective
effects into account. When available, radial velocities 
were taken from SIMBAD \citep{2000A&AS..143....9W}; otherwise, a value of zero was used. 
The VLBI data were assumed to be uncorrelated, that is,\ all matrices $\vec{V}_i$ were taken as diagonal.
For the present application, the most critical correlation is that between position and proper motion, 
which should normally be small if $t$ is close to the mean epoch of the VLBI measurements. 

\subsection{\textit{Gaia} data}
\label{sec:gaia}

\textit{Gaia} DR2 identifiers for the optical counterparts of the radio sources are given 
in Table~\ref{table2}. For most sources they were taken directly from SIMBAD, but in a few cases
(HD~22468, T~Tau, VY~CMa, and AR~Lac), they had to be derived by cross-matching the
radio positions with the \textit{Gaia} positions. Relevant optical data were retrieved from 
the \textit{Gaia} Archive\footnote{\url{https://gea.esac.esa.int/archive/}} 
and the full covariance matrices $\vec{C}_i$ computed from the formal uncertainties and 
correlation coefficients. As these data are readily available on-line, they are not reproduced 
in Table~\ref{table2}, with the exception of the $G$ magnitude and the re-normalised unit 
weight error (RUWE). The latter, computed from Archive data as described in \citet{LL:LL-124},
is a goodness-of-fit measure (formally the square root of the reduced chi-square of the astrometric 
solution) that 
should be around 1.0 for an astrometrically well-behaved source. $\text{RUWE}\gtrsim 1.4$ could 
indicate an astrometric binary, a (partially) resolved binary or multiple star, or an otherwise problematic 
source. The remaining columns in Table~\ref{table2} are explained below.

It is known that the parallaxes in \textit{Gaia} DR2 are systematically too small by a few tens of
$\text{microarcseconds}$ \citep{2018A&A...616A..17A}. The zero-point is estimated to be about $-0.03$~mas
for the faint quasars \citep{2018A&A...616A...2L}, but there is strong evidence that the bright stars 
of interest here have a more negative zero-point, around $-0.05$~mas
\citep[e.g.][]{2018ApJ...861..126R,2018arXiv180502650Z,2019arXiv190202355S}. This
is broadly confirmed by the VLBI data: the median parallax difference for the 41 radio stars in 
Table~\ref{table1} is $\varpi_\text{DR2}-\varpi_\text{VLBI} = -0.076 \pm 0.025$~mas. In the 
solutions reported below, the \textit{Gaia} DR2 parallaxes of all the radio stars have been increased
by 0.05~mas to take the zero-point error into account. Because parallax
and the other astrometric parameters in the \textit{Gaia} data are correlated, this changes the estimated rotation 
parameters, but only by small amounts: about 0.010~mas in $\vec{\varepsilon}(2015.5)$ 
and 0.005~mas~yr$^{-1}$ in $\vec{\omega}$.

\begin{figure*}[t]
\centerline{\resizebox{18cm}{!}{\includegraphics{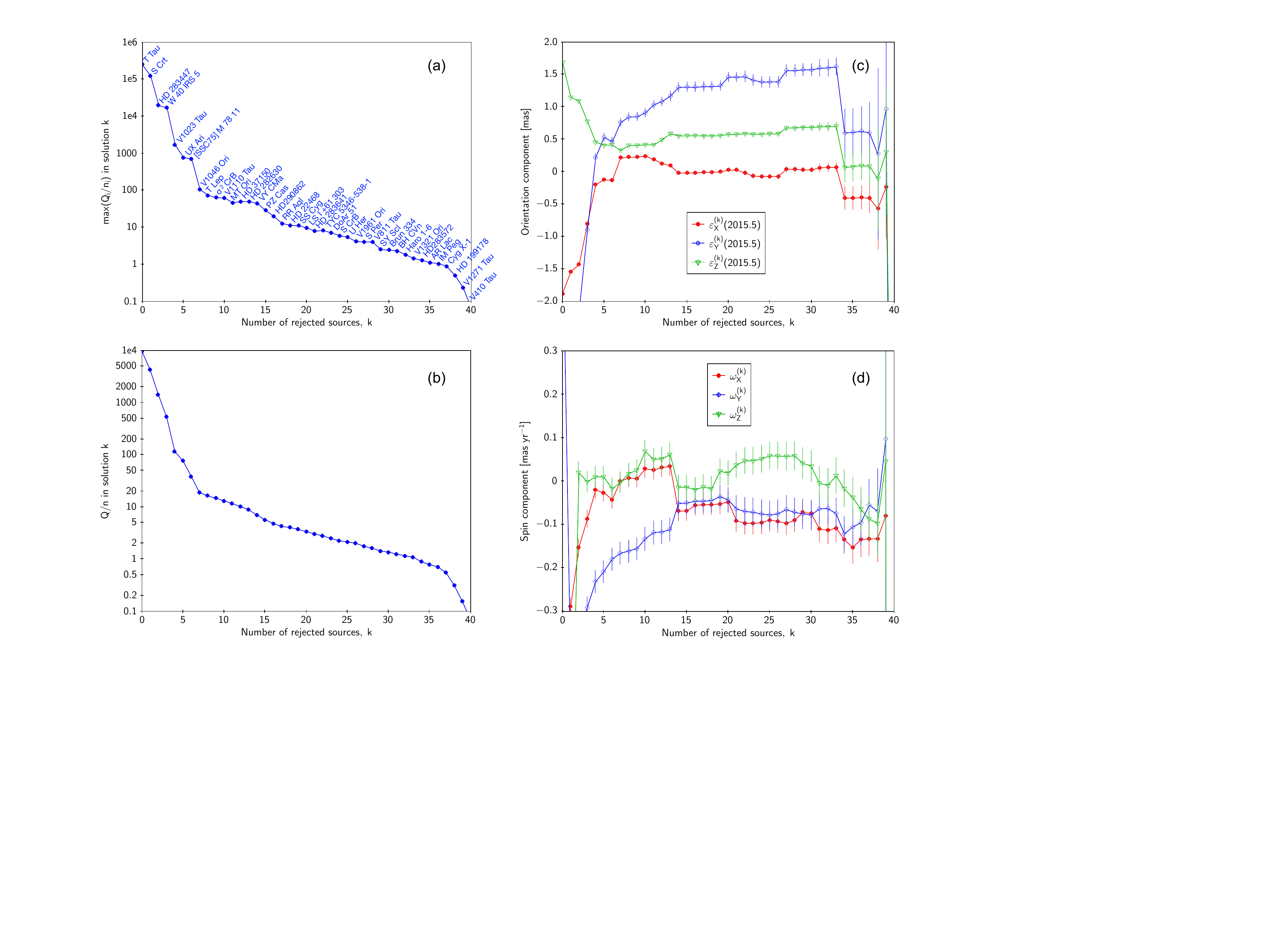}}}
\caption{Evolution of discrepancy measures and rotation parameters in a series of solutions 
where the $k$ most discrepant sources have been removed. 
\textbf{a)} Discrepancy measure $Q_i/n_i$ for the most discrepant source in the solution. 
Points are labelled with the name of the source.
\textbf{b)} Total discrepancy measure $Q/n$ as a function of the number of rejected
sources.
\textbf{c)} Estimated orientation parameters and \textbf{d)} spin parameters as functions 
of the number of rejected sources.
Error bars are formal $\pm 1\sigma$ uncertainties from the inverse normal matrix, 
i.e.\ not adjusted depending on the goodness-of-fit.\label{fig34}}
\end{figure*}

\begin{table*}
\caption{Summary of the different solutions for the orientation and spin parameters.}
\centering
\footnotesize
\begin{tabular}{ccccccc}
\hline\hline\noalign{\medskip}
Solution & \multicolumn{3}{c}{Orientation at $T=\text{J2015.5}$ (mas)}
& \multicolumn{3}{c}{Spin (mas~yr$^{-1}$)}\\
& $\varepsilon_X(T)$ & $\varepsilon_Y(T)$ & $\varepsilon_Z(T)$ 
& $\omega_X$ & $\omega_Y$ & $\omega_Z$ \\
\noalign{\smallskip}\hline\noalign{\medskip}
A & $-0.019\pm 0.032$ & $+1.304\pm 0.074$ & $+0.553\pm 0.026$
& $-0.068\pm 0.023$ & $-0.051\pm 0.027$ & $-0.014\pm 0.028$\\
B & -- & -- & --
& $-0.050\pm 0.036$ & $-0.139\pm 0.055$ & $+0.002\pm 0.038$\\
C & $-0.001\pm 0.036$ & $+1.413\pm 0.085$ & $+0.559\pm 0.033$
& $-0.056\pm 0.028$ & $-0.057\pm 0.028$ & $-0.019\pm 0.035$\\
D & $+0.026\pm 0.034$ & $+1.465\pm 0.079$ & $+0.629\pm 0.028$
& $-0.051\pm 0.025$ & $-0.066\pm 0.029$ & $-0.015\pm 0.030$\\
\noalign{\smallskip}\hline\noalign{\smallskip}
Adopted & $-0.019\pm 0.158$ & $+1.304\pm 0.349$ & $+0.553\pm 0.135$
& $-0.068\pm 0.052$ & $-0.051\pm 0.045$ & $-0.014\pm 0.066$\\
\noalign{\smallskip}\hline
\end{tabular}\label{table3}
\tablefoot{The table gives the estimated orientation and spin components for different solutions
with their uncertainties (after $\pm$). All solutions use the same subset of 26 sources, denoted as accepted
in Table~\ref{table2}. Solution~A is the baseline solution described in Sect.~\ref{sec:result}.
Solutions~B--D are the alternative solutions described in Sect.~\ref{sec:alt}: B using only 
proper motions, C using only positions, and D using the same data as A and the magnitude dependent 
model in Eq.~(\ref{eq:est20}). The `Adopted' solution is identical to A, but the uncertainties
are estimated by bootstrap resampling of the 26 sources (while for A--D the uncertainties are formal 
values computed from the inverse normal matrix).}
\end{table*}

\subsection{Results}
\label{sec:result}

The direct application of the solution method in Sect.~2.3 to the full sample of 41 sources 
gives a very poor fit as measured by the loss function $Q\simeq 2\,176\,000$ for 
$n\equiv\sum_i n_i=224$ degrees of freedom, or a reduced chi-square of
$Q/n\simeq 9713$. This solution also gives unrealistically large values for the spin parameters, 
where $|\vec{\hat{\omega}}|\simeq 2.5$~mas~yr$^{-1}$. Inspection of the discrepancy measure
$Q_i/n_i$ of the individual sources shows that T~Tau has by far the highest value at 
$Q_i/n_i\simeq 247\,502$, followed by S~Crt at $Q_i/n_i\simeq 123\,397$, and so on. 
Removing T~Tau from the sample and re-computing the solution and discrepancy measures
gives $Q\simeq 924\,169$ for $n=\sum_i n_i = 219$ degrees of freedom ($Q/n\simeq 4220$). In this solution
the source with the largest discrepancy measure is S~Crt at $Q_i/n_i\simeq 123\,714$.
Removing this source as well and iterating the procedure until all sources but one have been
removed gives a series of solutions with $k=0,~1,~\dots$ rejected sources. The evolution
of $\max(Q_i/n_i)$ and $Q/n$ as functions of $k$ is shown in the left panels of 
Fig.~\ref{fig34}; the corresponding orientation and spin parameters are shown in the right 
panels.  

Errors in the rotation parameters caused by non-linear source motions and other model 
deficiencies are generally reduced as more outliers are removed, while the statistical (formal)
uncertainties increase because fewer sources contribute to the solution. The 
optimum solution is a compromise between the opposite tendencies and may be found 
somewhere along the sequence of solutions described above. The rather smooth progression 
of the discrepancy measures in Figs.~\ref{fig34}a and b gives no clear indication of the optimum 
$k$, except that it is probably in the range from 7 (removing the seven most obvious outliers) 
to $\simeq\,$33 (after which $Q/n<1$). The spin parameters in Fig.~\ref{fig34}d show 
an abrupt change with the removal of HD~282630 at $k=13$, after which the fluctuations,
although not negligible, are roughly of a size that is compatible with the formal uncertainties.
For the subsequent discussion we assume, somewhat arbitrarily, the solution at $k=15$ as the 
baseline.%
\footnote{This choice of $k$ is the same as made in the original version of this paper.
Although that choice, based on an incorrect computation, is in principle irrelevant for the 
present analysis, the corrected results do not present a clear case for adopting
a different $k$. As it happens, $k=15$ results in exactly the same subset of rejected radio 
sources as in the original paper.}
This solution has $Q/n=5.68$ with $n=139$ degrees of freedom and the 
most discrepant source is PZ~Cas with $Q_i/n_i\simeq 28.604$. 

The accepted and rejected sources and their individual discrepancy measures for the
baseline solution are listed in Table~\ref{table2}. The rotation parameters are given in 
Table~\ref{table3} under entry A, including the formal uncertainties calculated from the
inverse normal matrix. The correlation matrix (for $T=\text{J2015.5}$) is
\setcounter{equation}{29}
\begin{multline}\label{eq:corrA}
\text{corr}\,\Bigl[\varepsilon_X(T),~\varepsilon_Y(T),~\varepsilon_Z(T),
~\omega_X,~\omega_Y,~\omega_Z\Bigr]\\
=\begin{bmatrix}
+1.000\! & +0.517\! & +0.204\! & +0.136\! & +0.066\! & +0.120\\
+0.517\! & +1.000\! & +0.206\! & +0.099\! & +0.131\! & +0.105\\
+0.204\! & +0.206\! & +1.000\! & +0.091\! & +0.035\! & +0.133\\
+0.136\! & +0.099\! & +0.091\! & +1.000\! & -0.028\! & +0.405\\
+0.066\! & +0.131\! & +0.035\! & -0.028\! & +1.000\! & -0.090\\
+0.120\! & +0.105\! & +0.133\! & +0.405\! & -0.090\! & +1.000\\
\end{bmatrix}.
\end{multline}

Given the reduced chi-square of the solution, $Q/n\simeq 5.68$,
it is likely that the formal uncertainties underestimate the actual errors. More
realistic estimates may be obtained by bootstrap resampling of the 26 accepted sources,
yielding the uncertainties in the `Adopted' entry of Table~\ref{table3}.

\begin{figure}[t]
\centerline{\resizebox{0.85\hsize}{!}{\includegraphics{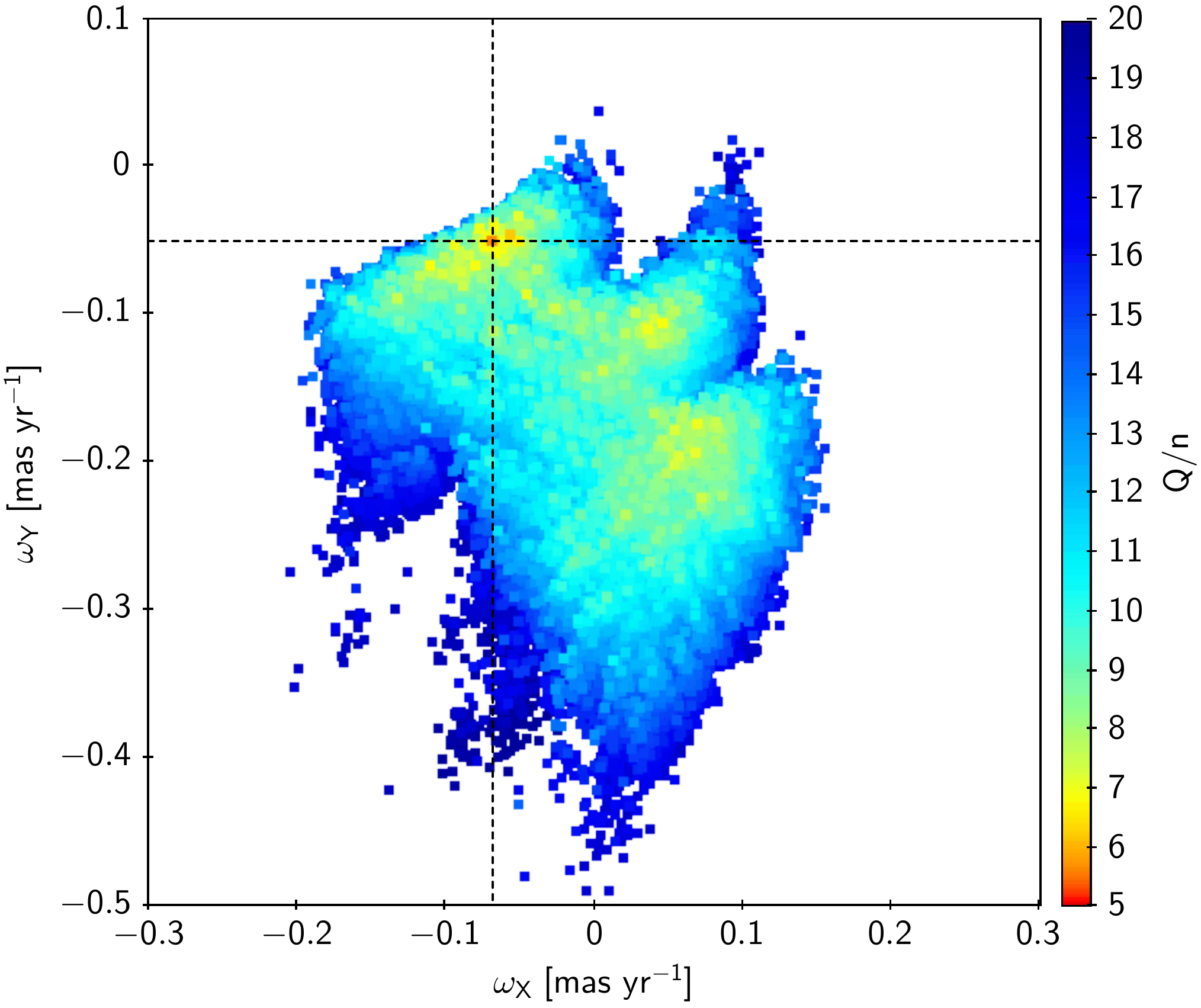}}}
\centerline{\resizebox{0.85\hsize}{!}{\includegraphics{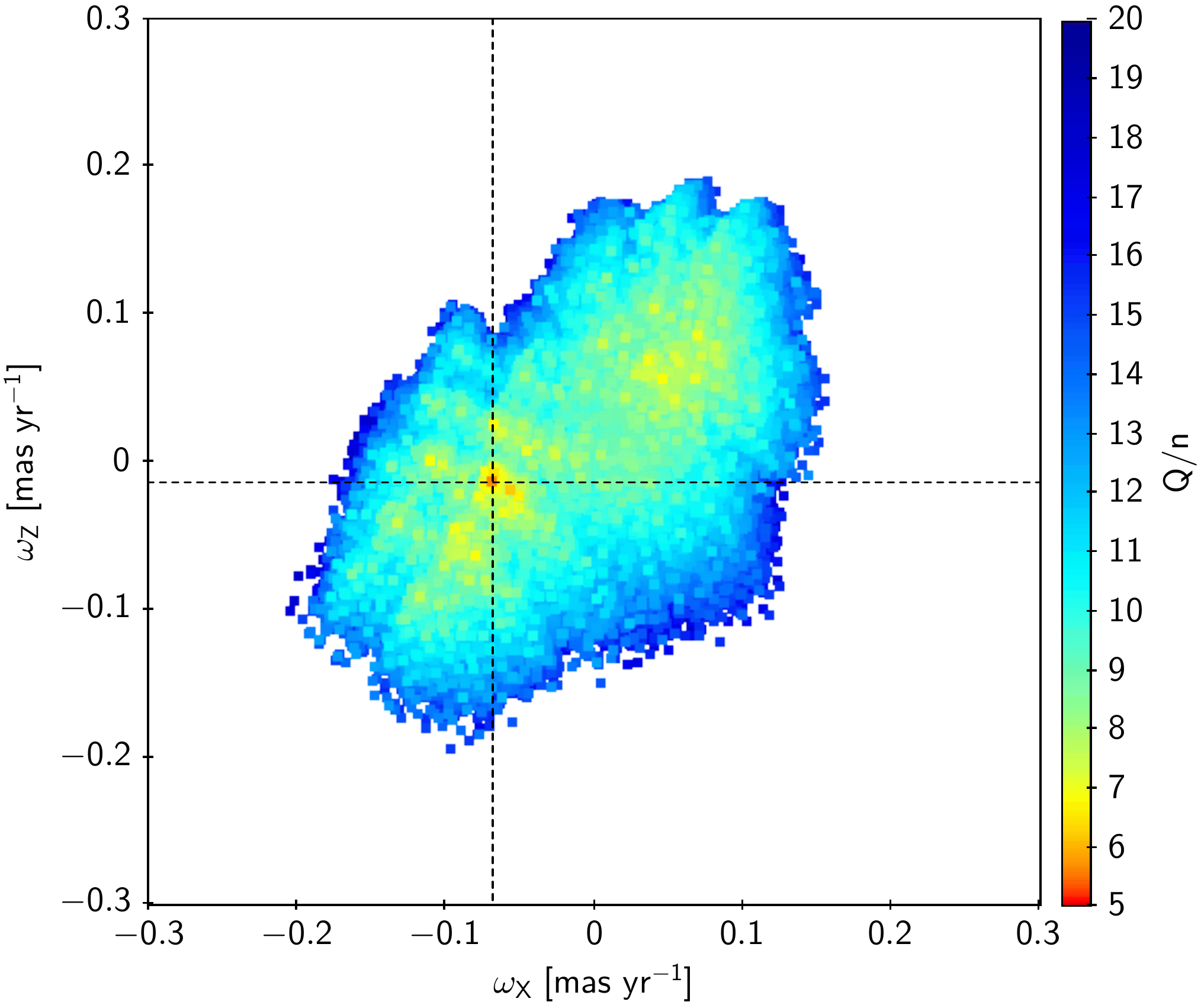}}}
\centerline{\resizebox{0.85\hsize}{!}{\includegraphics{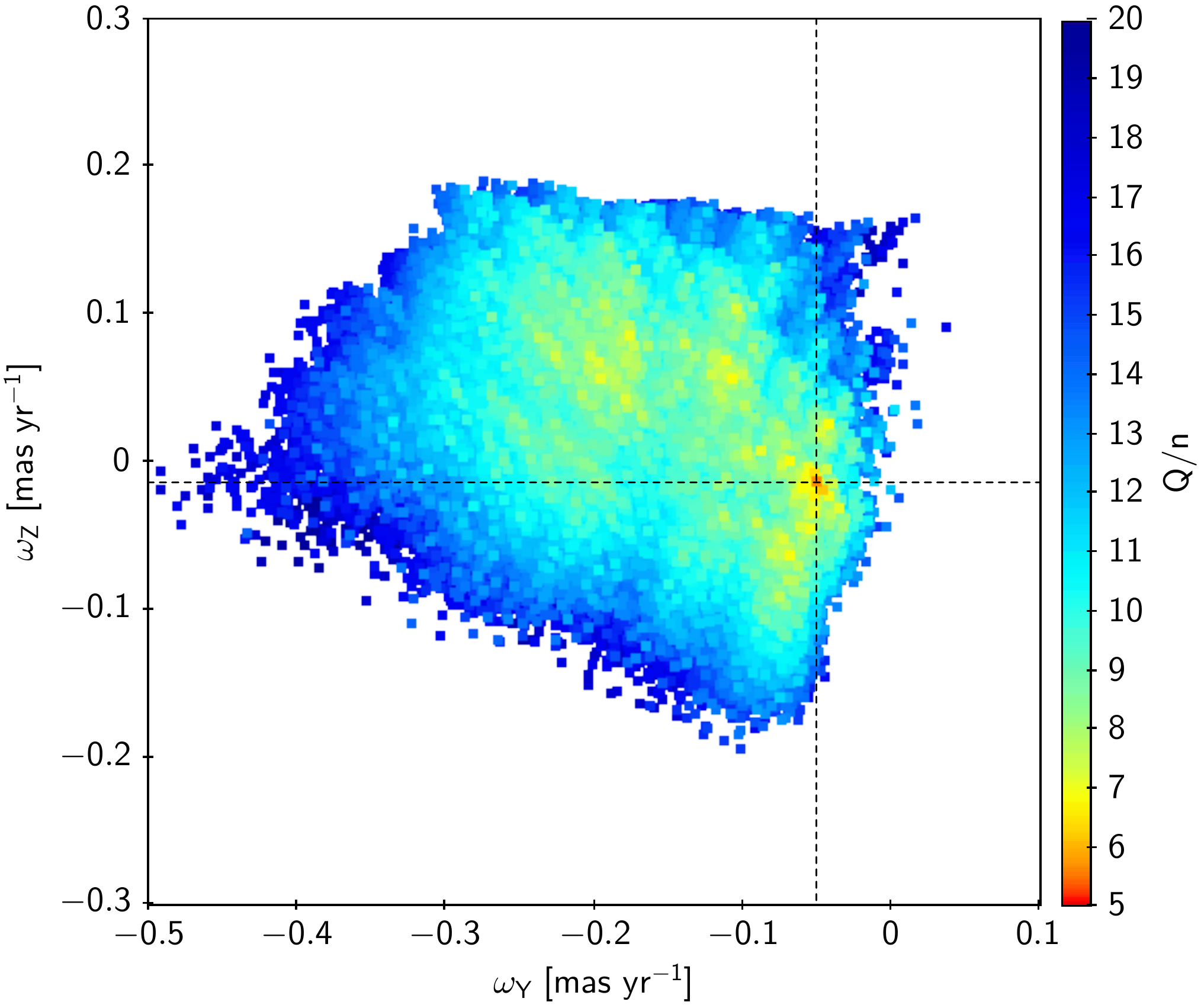}}}
\caption{Sensitivity of the solution to the selection of sources. The plots show the spin parameters 
and associated $Q/n$ values for $\text{about four}$~million solutions using different subsets of the sources 
as described in the text. The colour shows the smallest $Q/n$ in each bin. The
dashed lines denote the spin parameters for the baseline solution A with $Q/n\simeq 5.68$. \label{fig05}}
\end{figure}

From the original sample of $m=41$ sources, $k=15$ were thus removed to obtain 
the baseline subset of $m-k=26$ sources. It is not obvious that the process of successively 
removing the most discrepant source leads to the optimum subset in the sense 
that no other subset of the same size has a smaller $Q/n$. It is conceivable that a different 
procedure, for instance,\ starting from a smaller subset and adding the best-fitting sources 
\citep[outward selection;][]{Ben-Gal2010}, would lead to a different result. While it is 
impractical to test all $\binom{41}{26}\simeq 6\times 10^{10}$ possible combinations, an 
exhaustive search of the $\binom{33}{26}=4\,272\,048$ different subsets of size 26 drawn 
from the 33 sources with the smallest discrepancy measure did not uncover a
more favourable subset. Figure~\ref{fig05} shows $Q/n$ versus the spin components
for these solutions. Many of these solutions are very different from the baseline solution in
terms of the spin components, but invariably their $Q/n$ is then also significantly higher.
This makes it credible that the spin parameters of the adopted solution are not the 
chance result of a particular combination of data for a few sources.

The solution gives improved estimates of the astrometric parameters of the sources,
obtained by solving (21) for each $i$. These results are not tabulated, as they 
are practically the same as a weighted average of the data obtained by
very long baseline interferometry (VLBI) and the 
\textit{Gaia} data, after correcting the latter for the frame rotation and parallax zero-point. 
For example, the joint estimate of the parallax of Cyg~X-1 is $\hat{\varpi}=0.504\pm 0.025$~mas, 
which is very close to the weighted mean of the VLBI value, $0.547\pm 0.041$~mas, and the \textit{Gaia}
DR2 value after correction for the zero-point, $0.472\pm 0.032$~mas.

\subsection{Alternative solutions}
\label{sec:alt}

The classical way to determine $\vec{\omega}$ is to solve the over-determined system of
Eqs.~(8) and (9) using only the proper motion differences. In this process it is natural 
to assign a weight to each equation that is inversely proportional to the combined variances of the 
VLBI and \textit{Gaia} proper motions. An equivalently weighted least-squares solution is obtained 
with the formalism of Sect.~2.3 simply by deleting in $\vec{f}_i$ all the VLBI data items
that are not proper motions.
The resulting normal equations, Eq.~(22), are of course singular for the orientation parameters,
but the lower right $3\times 3$ part of the equations gives the desired solution for $\vec{\omega}$.
Applying this procedure to the baseline subset of 26 sources gives the result shown as 
solution B in Table~\ref{table3}. The spin parameters are reasonably close to those of the baseline 
solution (A) for the $X$ and $Z$ components, while the $Y$ component shows a more negative value. 
It is more interesting, 
however, that the formal uncertainties are significantly higher in solution B than in A: the uncertainty
is a factor two higher for the $Y$ component.

It thus appears that the positional VLBI data are at least as valuable as the proper motion data
when estimating the spin, at least for the spread of VLBI epochs considered in this work. A direct test of 
this hypothesis is to make the complementary solution to what is described above, that is,\ deleting 
all proper motion items in $\vec{f}_i$. The result is shown as solution C in Table~\ref{table3}. 
In this case, the spin is more precise (in terms of formal uncertainties) than in solution B where 
(only) the proper motions  are used, although not as precise as in A. 
It can be noted that only 23 of the 26 sources contribute to 
solution C because no positional VLBI data are provided for S~CrB, U~Her, and RR~Aql,
although these sources belong to the baseline subset.

Figure~4 of \citet{2018A&A...616A...2L} suggests that the transition from the faint to the
bright reference frame in \textit{Gaia} DR2 does not occur abruptly at $G=13,$ but
happens gradually from $G\simeq 13$ to $\simeq\! 11$~mag. Several of the sources in Table~1
have magnitudes in the transition interval and may therefore not contribute fully to the
determination of the rotation parameters. 
In solution D the model of the \textit{Gaia} data in (11) is modified so that the 
applied rotation is $\vec{x}$ multiplied by the magnitude-dependent function
\begin{equation}\label{eq:est20}
\phi(G) = \begin{cases} 1 &\text{if $G\le 11$,}\\ 
(13-G)/2 &\text{if $11< G \le 13$,}\\  0 &\text{if $G>13$.} \end{cases}
\end{equation}  
This improves the overall fit significantly ($Q=747.6$ in D against 789.5 in A), although the 
discrepancy measure increases for two of the six accepted sources in the magnitude range 
11--13 (V1961~Ori and DoAr~51). The VLBI data therefore support the magnitude dependent 
model, although not unambiguously. The resulting rotation parameters (for 
$G\lesssim 11$) are not significantly different from the baseline solution.

\subsection{Weights of the individual sources}
\label{sec:weight}

Table~\ref{table2} includes the statistics $E_i$ and $\Omega_i$ from Eq.~(\ref{est15}),
indicating the potential weights of the sources in the orientation and spin solutions.
The sources that contribute most weight to the spin solution (largest $\Omega_i$)
are AR~Lac, LS~I~+61~303, Cyg~X-1, HD~199178, and V410~Tau; all of them
have $\Omega_i>300$~mas$^{-2}$~yr$^2$ and all are included in the 
baseline subset. The first three and HD~199178 mainly contribute by virtue of their relatively
small positional uncertainties ($\simeq 0.3$~mas) at a very early epoch, $t\simeq 1992$.
IM~Peg, one of the best-observed radio stars \citep{2015CQGra..32v4021B}, has a much 
smaller weight in this analysis because its uncertainties in \textit{Gaia} DR2 are relatively 
large.

The main contributions to the determination of the orientation (largest $E_i$) come from
young stellar objects in the Taurus and Orion regions that are observed as part of the GOBELINS
survey \citep{2018ApJ...859...33G,2017ApJ...834..142K}, through their high precision 
and proximity in time with the \textit{Gaia} DR2 epoch. The strong concentration of these 
sources in a small part of the sky, approximately in the $+Y$ direction, is responsible
for the relatively large uncertainty of the $Y$ component of $\vec{\varepsilon}(2015.5)$
in Table~\ref{table3}.

\begin{figure}[t]
\centerline{\resizebox{0.9\hsize}{!}{\includegraphics{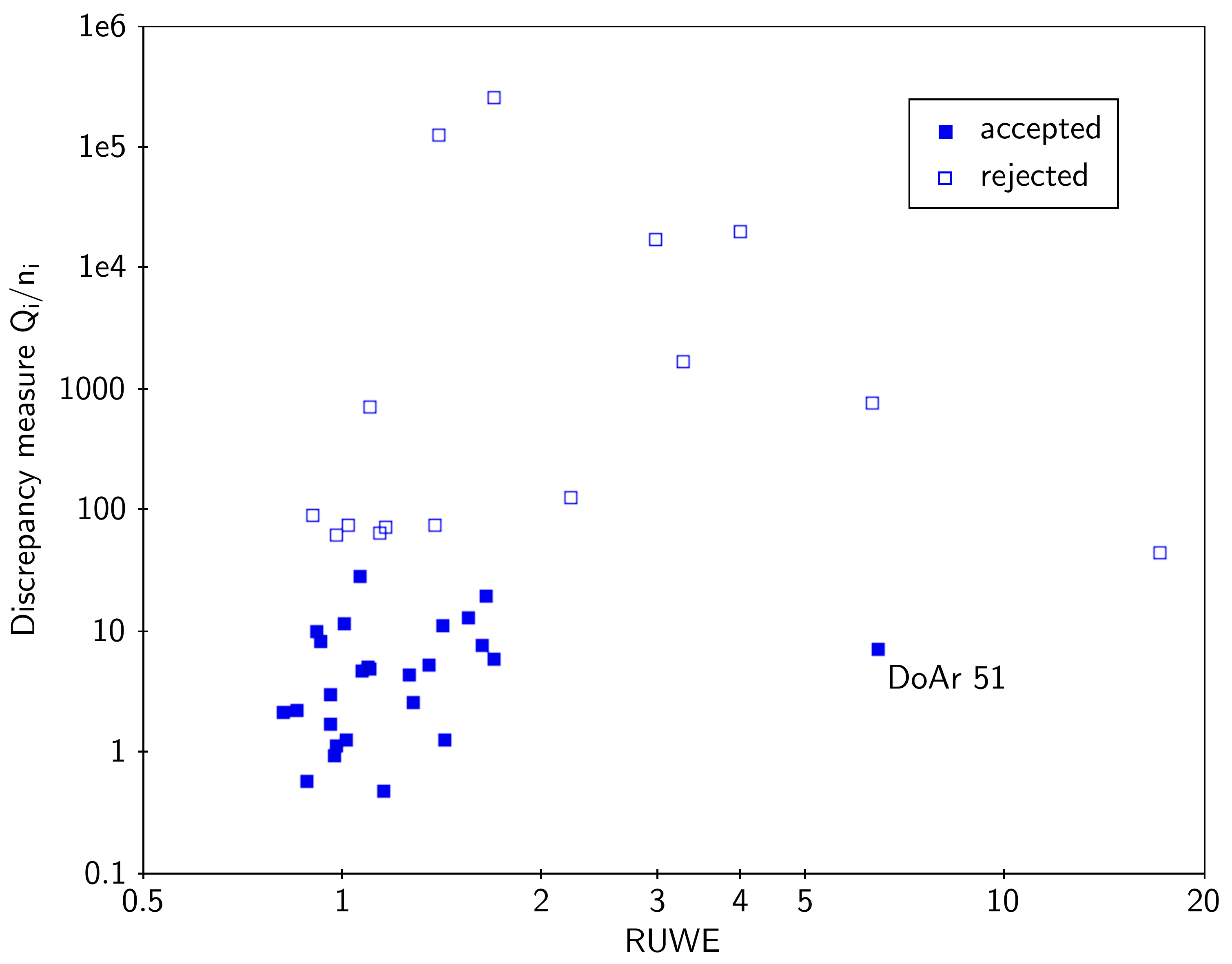}}}
\caption{Discrepancy measures of the sources in solution A against the RUWE in the 
\textit{Gaia} DR2 astrometry. The case of DoAr~51 is discussed in Sect.~\ref{sec:discBin}.\label{fig06}}
\end{figure}

\section{Discussion}
\label{sec:disc}

\subsection{Estimated rotation parameters}
\label{sec:discRot}

The results in Table~\ref{table3} yield a spin for the \textit{Gaia} DR2 bright reference frame
of $\vec{\omega}=(-0.068,-0.051,-0.014)$~mas~yr$^{-1}$, which is not statistically significant 
in view of the uncertainties obtained by the bootstrap method. Nevertheless, the consistently 
negative values of $\omega_X$ and $\omega_Y$ obtained for a wide range of $k$ in 
Fig.~\ref{fig34}d do suggest a marginally significant spin of the order of 0.1~mas~yr$^{-1}$.
This is supported by the qualitative agreement between solutions B and C, which use 
different subsets of the VLBI data.

With regard to the orientation parameters $\vec{\varepsilon}(2015.5)$, the $Y$ and $Z$ 
components are also statistically significant  in view of the uncertainties obtained by the 
bootstrap method and indicate a total orientation error of about 1.4~mas.
The large uncertainty of the $Y$ component reflects the unfavourable celestial
distribution of the more recent VLBI observations included in the analysis 
(Sect.~3.5).

As mentioned in the Introduction, the non-zero spin of the bright reference frame of \textit{Gaia} DR2
is also seen in a comparison with proper motions of \textsc{Hipparcos} stars calculated from 
the position differences between \textit{Gaia} DR2 and the \textsc{Hipparcos} catalogue, 
divided by the $\sim$24~yr epoch difference. This comparison was made by 
\citet{2018ApJS..239...31B} in the course of 
constructing \textit{The \textit{Hipparcos}--\textit{Gaia} Catalog of Accelerations} (HGCA). 
For 115\,663 \textsc{Hipparcos} stars the HGCA gives three essentially independent sets 
of proper motions, namely (i) as measured by \textsc{Hipparcos} around epoch 1991.25,
(ii) as measured by \textit{Gaia} (in DR2) around epoch 2015.5, and (iii) as calculated from the
\textsc{Hipparcos}--\textit{Gaia} position differences divided by the epoch difference. 
The HGCA is intended for orbit fitting and for identifying candidate stars with substellar or 
dark companions. To this end, \citet{2018ApJS..239...31B} made a careful cross-calibration 
of the three sets of proper motions in order to eliminate any systematic offsets, in particular 
the global rotations. For the rotation between sets (ii) and (iii) the result as given in Brandt's Table~1
is $\vec{\omega}=(-0.081,-0.113,-0.038)$~mas~yr$^{-1}$. Brandt used the same sign
convention for this vector as in this paper, so that his result can be directly compared with
our Table~\ref{table3}. 

The comparison with \citet{2018ApJS..239...31B} rests on the assumption that the 
\textit{Hipparcos}--\textit{Gaia} position differences yield proper motions that are 
absolute; that is, they are expressed in a reference frame that is non-rotating with respect
to the International Celestial Reference System (ICRS).%
\footnote{\citet{2018ApJS..239...31B} adopted the (bright) reference 
frame of \textit{Gaia} DR2 for the HGCA, therefore the published proper motions of type (iii),
called \texttt{pmra\_hg} and \texttt{pmdec\_hg} in his Table~5, are not absolute 
in our sense. To place these values on the ICRS, the applied
cross-calibration corrections must be subtracted, i.e.\ use \texttt{pmra\_hg}--\texttt{crosscal\_pmra\_hg} 
and \texttt{pmdec\_hg}--\texttt{crosscal\_pmdec\_hg}.}
The validity of this assumption depends on the quality of the positional reference
frame of \textsc{Hipparcos} at epoch 1991.25 and of the positional reference frame for the 
same stars in \textit{Gaia} DR2 at epoch 2015.5. If both sets of positions are aligned 
with the ICRS at their respective epochs, the proper motions calculated from the position 
differences must clearly be non-rotating with respect to the ICRS. 

According to the present analysis, the bright reference frame of \text{Gaia} DR2 
was however offset from ICRS by more than 1~mas at epoch 2015.5, as given by 
$\vec{\varepsilon}(2015.5)$ in Table~\ref{table3}. To take this into 
account in the spin comparison, Brandt's estimate quoted above needs to be increased by 
$\vec{\varepsilon}(2015.5)/(24.25~\text{yr})$. This yields the corrected estimate of 
$\vec{\omega}=(-0.082,-0.059,-0.015)$~mas~yr$^{-1}$, which is in good agreement with the adopted spin solution in Table~\ref{table3}. The uncertainty of this estimate is
$\sqrt{0.6^2+0.4^2}/24.25\simeq 0.03$~mas~yr$^{-1}$ per axis, which was obtained by combining
the RMS orientation error of the \textsc{Hipparcos} reference frame at epoch 1991.25 
\citep[in Vol.~3, Sect.~18.7 of \textit{The Hipparcos and Tycho Catalogues},][given as 
0.6~mas per axis]{hip:catalogue} and the uncertainty of $\vec{\varepsilon}(2015.5)$, 
in this work taken to be 0.4~mas.

The old \textsc{Hipparcos} positions thus provide an independent estimate of the spin, 
which agrees with with the results from the VLBI observations and is of comparable or 
even better accuracy. However, because its uncertainty is dominated by the unknown orientation 
errors of the \textsc{Hipparcos} reference frame, which will not improve, the method will 
be of limited value for the validation of future \textit{Gaia} data releases.
In contrast, the VLBI method, as discussed in Sect.~4.4, has great potential 
for future improvement.

The origin of the non-zero rotation parameters for the bright reference frame of 
\textit{Gaia} DR2 is a deficiency in the specific instrument calibration model used for the DR2 
astrometric solution. The relevance of this deficiency for the reference frame was not recognised at the time when the DR2 data were prepared and validated,
and although its effect on the bright sources was noted 
\citep[e.g.\ Fig.~4 in][]{2018A&A...616A...2L}, no explanation of its origin was offered. 
This is now understood, and the calibration model is being improved with a view towards avoiding 
a similar error in future data releases (see Appendix~B).

\subsection{Use of positional VLBI data}
\label{sec:discPos}

A comparison of the formal uncertainties of $\vec{\omega}$ in solutions A, B, and C 
(Table~\ref{table3}) demonstrates the advantage of including positional VLBI data
in a joint solution for the orientation and spin parameters. Not only do the positional 
data, as already noted, improve the spin solution (A is better than B for $\vec{\omega}$), 
but the inclusion of proper motion data also improves the orientation parameters
(A is better than C for $\vec{\varepsilon}$). 

The value of positional VLBI data for the spin is especially noteworthy because it means that 
even single-epoch VLBI astrometry can be incorporated in the solution. This will contribute 
to the spin determination, especially if the data are taken at an epoch that is well separated from the \textit{Gaia}
epoch. This is important to keep in mind, both for the inclusion of old, possibly unpublished
VLBI measurements (Sect.~\ref{sec:discPlea}) and for the scheduling of future VLBI sessions 
specifically for the reference frame (Sect.~\ref{sec:discFut}). The latter need not be constrained by 
considerations of parallax factor and temporal spread of the measurements for the proper motions.

\subsection{Binarity and source structure}
\label{sec:discBin}

About half of the radio sources in Table~\ref{table1} are known to be binaries or members 
of double or multiple systems. Interacting pairs with periods from a day to tens of days
include the RS CVn systems and high-mass X-ray binaries, which usually provide good fits
to the single-star model unless they are perturbed by a more distant component, as is the case 
for $\sigma^2$~CrB. Binaries with periods of years to hundreds of years are more problematic
unless a complete orbit can be determined. Orbits have been determined for some radio 
stars, but as the corresponding binary data from \textit{Gaia} are not yet available, they are
not included in this analysis. For other objects, the VLBI observations have detected the 
curvature of a long-period orbit by means of acceleration terms; this is the case for example\ for
UX~Ari, HD~283447, and T~Tau, all of which obtain large discrepancy measures in the present 
analysis (Fig.~\ref{fig34}a).

The radio stars included in the present analysis have not been a~priori screened for known 
or anticipated problems with multiplicity and source structure. Except for known binaries, 
several Mira variables and red supergiants have therefore been included, although they are far from ideal 
targets because of the issues described in Sect.~\ref{sec:vlbi}. It is very likely that the optical 
\textit{Gaia} observations are also severely affected by the extended and complex atmospheres
\citep{2011A&A...528A.120C}. This type of radio star should probably be avoided entirely
for the reference frame.  

A statistic that could be used for screening the sources is the RUWE given in Table~\ref{table2}. The RUWE measures how well the different \textit{Gaia} 
observations that were made over a few years agree with the five-parameter single-star model. It is therefore mainly sensitive to the presence of companions at separations from a few milliarcsecond to about 
one arcsecond, and a relevant indicator of potentially problematic sources of that particular kind.
It is not sensitive to non-linear motions, if the deviation from linear is small over the few years of 
the \textit{Gaia} observations. A large discrepancy measure $Q_i/n_i$ also indicates a 
problematic source, but of a rather different kind. In contrast to the RUWE, it is sensitive 
to radio--optical offsets, and it is also more sensitive to long-period perturbations if the
VLBI and \textit{Gaia} measurements are made at very different epochs.
Figure~\ref{fig06} shows a weak positive correlation between the two statistics, which is 
to be expected because their different regimes of sensitivity overlap. 
However, it is clear that RUWE alone is not sufficient to find the best candidate targets for
determination of the frame rotation parameters. 

In Fig.~\ref{fig06} the source DoAr~51 stands out because it has a large $\text{RUWE}=6.43$, 
while obtaining a reasonable fit in the solution ($Q_i/n_i=7.127$). This object is a 
triple system consisting of an equal-mass pair with a period of about eight~years, a separation 
of approximately 56~mas at 2015.5, and a fainter 
tertiary component at a separation of about 790~mas \citep{2018AJ....155..109S}. This
triple configuration could explain the high RUWE obtained with \textit{Gaia}. The VLBI
observations detect both components of the close pair, and the data in Table~1
refer to their centre of mass \citep{2017ApJ...834..141O}. The near-coincidence of the 
VLBI and \textit{Gaia} epochs, together with the near-coincidence of the optical photocentre with 
the centre of mass of the close pair and the rather large uncertainty of the \textit{Gaia} proper 
motion ($\sim$1~mas~yr$^{-1}$), could explain why the discrepancy measure is not higher. 
With $\Omega_i=3.2$~mas$^{-2}$~yr$^2$, this object contributes less to the determination 
of $\vec{\omega}$ than any of the other accepted sources in Table~\ref{table2}. DoAr~51 
is a good example of an object for which an extended 
model along the lines in Sect.~2.4 could drastically increase the usefulness of the data.


\subsection{Precision of future solutions}
\label{sec:discFut}

The bright reference frame in future releases of \textit{Gaia} data should ideally be validated 
at a level compatible with the expected errors of the proper motions, which may be as small
as a few $\mu$as~yr$^{-1}$. This will require a much better accuracy in the spin parameters 
than is achieved in the present analysis. Clearly, the uncertainties can be reduced by 
including VLBI data for more radio stars, and/or using improved data for the sources already 
considered. As discussed in Sect.~\ref{sec:discPos}, the epoch of the added data is an
important factor. If dedicated VLBI measurements are contemplated for this purpose, different 
scenarios can be envisaged concerning the number, distribution, and epochs of the planned
observations, and it is of great interest to predict the accuracy that can be achieved in various 
cases. This can be done by applying the algorithm in Sect.~\ref{sec:est} to simulated data sets,
using the left-hand side of Eq.~(\ref{est08}) to compute the formal precisions.

A major uncertainty in any such prediction is the extent to which binarity, source structure, and 
radio-optical offsets will limit the achievable accuracy. One extreme scenario is that these effects 
already dominate the error budget in the current analysis. Support for this could be drawn
from the fact that $Q_i/n_i>1$ for nearly all the accepted sources in Table~\ref{table2}.
In this scenario it will not help much to add more and better VLBI data for the radio stars already 
considered, and the safest way to improve the solution may be to increase the 
number of sources, $m$, and rely on the statistical improvement by $m^{-1/2}$.

\begin{table}
\caption{Formal improvement in the determination of orientation and spin parameters
expected from the addition of new positional VLBI data for the 26 sources in 
the baseline subsample.}
\centering
\footnotesize
\begin{tabular}{lccc}
\hline\hline\noalign{\medskip}
\textit{Gaia} data & Added VLBI data & $\sigma[\vec{\varepsilon}(T)]$ & $\sigma[\vec{\omega}]$ \\
& ($\sigma_\text{pos}=0.1$~mas) & ($\mu$as) & ($\mu\text{as~yr}^{-1}$)\\ 
\noalign{\smallskip}\hline\noalign{\medskip}
DR2 ($T=2015.5$) & none & 48.9 & 26.1\\
\multicolumn{1}{c}{\ditto} &  $t=2020$ & 25.3 & 21.8\\
\multicolumn{1}{c}{\ditto} &  $t=2025$ & 26.7 & 21.3\\
\multicolumn{1}{c}{\ditto} &  $t=2030$ & 27.8 & 21.2\\
\noalign{\smallskip}\hline\noalign{\smallskip}
5~yr ($T=2017.1$) & none & 49.6 & \phantom{0}9.6\\
\multicolumn{1}{c}{\ditto} & $t=2020$ & 22.1 & \phantom{0}7.0\\
\multicolumn{1}{c}{\ditto} & $t=2025$ & 22.9 & \phantom{0}5.8\\
\multicolumn{1}{c}{\ditto} & $t=2030$ & 23.7 & \phantom{0}5.3\\
\noalign{\smallskip}\hline\noalign{\smallskip}
10~yr ($T=2019.6$) & none & 57.0 & \phantom{0}6.9\\
\multicolumn{1}{c}{\ditto} & $t=2020$ & 22.5 & \phantom{0}4.6\\
\multicolumn{1}{c}{\ditto} & $t=2025$ & 21.0 & \phantom{0}3.6\\
\multicolumn{1}{c}{\ditto} & $t=2030$ & 21.7 & \phantom{0}2.9\\
\noalign{\smallskip}\hline
\end{tabular}\label{table4}
\tablefoot{$T$ is the reference epoch used in the solution; $t$ is the epoch of the added
VLBI observations. The last two columns give the formal uncertainties in $\vec{\varepsilon}(T)$
and $\vec{\omega}$, calculated as the quadratic means of the uncertainties in the 
$X$, $Y$, and $Z$ components. The top entry (DR2 without added VLBI data)  
corresponds to solution A in Table~\ref{table3}.}
\end{table}

Realistically, however, the prospects are not as bleak as outlined above. Better screening of
the sample, modelling of orbital motions and offsets, etc., will surely improve the results, and 
this process will be helped by the addition of new data both from VLBI and \textit{Gaia}. The 
scenario at the other extreme is that such improvements will allow us to reach the formal 
uncertainties computed from the least-squares equations. This (optimistic) assumption is
the basis for the predictions in Table~\ref{table4}.

The first entry in Table~\ref{table4} represents the baseline solution of the present
analysis (A in Table~\ref{table3}). For brevity, the formal uncertainties of the six rotation 
parameters are condensed here into two numbers, one for the orientation and one for the spin. 
For the next three entries, it is assumed that new VLBI observations of the same 26 accepted 
sources are obtained at the specified epoch $t$, with a precision of 0.1~mas in each coordinate.
Performing the analysis as in Sect.~\ref{sec:est} with the same \textit{Gaia} DR2 data,
we obtain the formal uncertainties in the last two columns. Remarkably, while the orientation 
(at $T=2015.5$) is better determined with the added data, the improvement is small in the 
spin and practically independent of the epoch of the added VLBI measurements.
This shows that the formal uncertainties of the spin in solution A are limited
by the accuracy of the proper motions in \textit{Gaia} DR2 rather than by the VLBI data.
The slight increase in $\sigma[\vec{\varepsilon}(T)]$ with $t$ arises because the effective
mean epoch of the (old plus new) VLBI measurements moves away from the reference
epoch $T$.

The middle four entries in Table~\ref{table4} show predictions when \textit{Gaia}
data are used based on the nominal mission length $L=5$~yr. It is assumed that the
uncertainties of the \textit{Gaia} astrometry improve as $L^{-1/2}$ for the positions and
parallaxes, and as $L^{-3/2}$ for the proper motions (taking $L=1.8$~yr for DR2). The
covariance matrices $\vec{C}_i$ in (\ref{est01}) are simply scaled by the corresponding 
factors, leaving the correlations unchanged from DR2. The orientation parameters now
refer to the corresponding reference epoch of the \textit{Gaia} observations, $T=2017.1$.%
\footnote{The reference epochs of future data releases are not known at the present time. 
The values in the table are assumed for the purpose of this study and correspond to the 
approximate intervals 2014.6--2016.4 (DR2), 2014.6--2019.6, and 2014.6--2024.6.}
Consistent with the previous finding that the DR2 precision is the main limitation 
in solution A, the improved \textit{Gaia} data drastically reduce the formal
uncertainty of the spin parameters even without any additional VLBI data. Adding new
data for the 26 stars improves the determination of the spin still further, especially if the 
new measurements are at a late epoch. 
The orientation parameters, on the other hand, are not much improved. Their
uncertainties are basically limited by the VLBI position errors, typically of the order
of 0.1~mas, and the small number of sources: $0.1/\!\sqrt{26}\simeq 0.02$~mas.  

The last four entries in Table~\ref{table4} are for an extended \textit{Gaia} mission
covering a full decade of observations ($L=10$~yr). Here the spin parameters receive 
another boost in precision, while there is almost no improvement in the orientation 
parameters.  The $\sigma[\vec{\varepsilon}(T)]$ has a shallow minimum at $t\simeq 2024$ 
because the effective mean epoch of the VLBI measurements is then close to the assumed
reference epoch, $T=2019.6$.  

Taking into account that the two extreme
scenarios considered above are probably to some extent true, we conclude from this crude assessment that a combination of actions 
should be taken to ensure that a robust and accurate estimate of the rotation parameters can be 
derived at the time when the final \textit{Gaia} results become available. These actions 
should include compiling and recalibrating past VLBI measurements, as well as securing new
observations of both old targets and as many new as possible.

\subsection{A plea to VLBI observers}
\label{sec:discPlea}

Most of the observational VLBI programmes used in this study address problems in Galactic or 
stellar astrophysics, and are primarily concerned with the parallaxes and proper motions
of the radio stars, or of their orbits, surface structures, and similar. Consequently, many of
the publications do not provide the (barycentric) positions that the authors must have derived
along with the parallaxes and proper motions by fitting an astrometric model to their positional
measurements. This is not a problem as long as the authors provide the individual measurements,
and their times, on which the fit was based. For this reason, as mentioned in Sect.~\ref{sec:vlbi}, 
several of the barycentric positions in Table~\ref{table1} were derived by the present author by
fitting the standard model (Appendix~\ref{sec:stand}) to the published VLBI measurements. 
Although this procedure is not without advantages (see below), it is of course simpler if the
full solution is provided by the observers.
  
In this context, it is useful to stress the unique historical value of positional data. 
In contrast to the parallaxes, for example, which can be re-determined at a future time, positional 
measurements can never be repeated, and for a number of applications their value only increases 
with time. A plea to observers making high-precision VLBI astrometry is therefore that they publish the 
full result of their astrometric fits, including the barycentric position and corresponding epoch. 

It is nevertheless good practice to publish the individual position measurements for future 
uses as well. This allows alternative models, which may\ include acceleration terms or orbital parameters  for example,
to be fitted in combination with other data. The general method described in Sect.~\ref{sec:est}  
is readily adapted to the use of individual VLBI measurements, and this has the advantage that
otherwise neglected correlations are fully taken into account. 

Relative astrometry using phase-referencing techniques are converted into absolute coordinates 
using an assumed position in the ICRS of the reference source (calibrator), usually a quasar. 
To first order,
small errors in the calibrator position directly transfer to the measured coordinates of the target 
\citep{2014ARA&A..52..339R}. This gives a constant offset of no consequence when the 
parallax and proper motion of the target are fitted to the data, 
but it might be important for the present application. 
It is customary 
to specify the calibrators used in phase-referencing observations, and sometimes also their
assumed positions. Knowing the identities and adopted positions of the calibrators is indeed
highly desirable because it allows the target positions to be corrected when improved 
calibrator positions become available (cf.\ Sect.~\ref{sec:vlbi}).

\section{Conclusions}
\label{sec:concl}

This paper provides a rigorous mathematical framework for estimating the orientation and 
spin of the \textit{Gaia} reference frame, in which the \textit{Gaia} data are optimally 
combined with VLBI measurements of bright radio sources. The simultaneous estimation
of the orientation ($\vec{\varepsilon}$) and spin ($\vec{\omega}$) is
essential for achieving the best accuracy. The method takes full 
advantage of past and future single-epoch VLBI measurements of \textit{Gaia} sources
for the determination of the spin. Independent estimates of their proper motions from 
VLBI can be incorporated into the solution, but are not required by the method.

Applied to published VLBI data for a sample of 41 bright ($G\le 13$~mag) radio sources,
the method gives the rotation parameters summarised in Table~\ref{table3}. The solution 
retains 26 of the investigated radio sources, while 15 are rejected based on a statistical 
discrepancy measure sensitive to binarity and source structure. The results indicate
that the bright reference frame of \textit{Gaia} DR2 at the reference epoch 2015.5 
is offset from the ICRS by about 1.3~mas in $Y$ and 0.6~mas in $Z$. The solution for 
the spin indicates a rotation rate of the order of 0.1~mas~yr$^{-1}$. The 
components of the spin solution are not statistically significant in comparison 
with their uncertainties from the bootstrap method. Nevertheless, they agree well 
with independent estimates obtained from a comparison of the extrapolated 
\textit{Gaia} DR2 positions with the \textsc{Hipparcos} catalogue at epoch J1991.25 
\citep{2018A&A...616A...2L,2018ApJS..239...31B} if the latter estimates are corrected
for the alignment error of \textit{Gaia} DR2 at the \textit{Gaia} epoch. 
The accuracy of the present study is limited by the relatively small number of radio stars included, 
by the uncertainties of the \textit{Gaia} DR2 proper motions, and by issues related to the 
astrophysical nature of the sources.

The origin of the spin of the bright reference frame of \textit{Gaia} DR2 is understood
and measures have been taken to to avoid this problem in future data releases
(see Appendix~B). 
Nevertheless, it is important that the consistency of future reference frames can be validated
across the full range of magnitudes and the present method offers such a possibility for 
the bright stars. As many as possible of the already existing VLBI measurements of radio 
stars should be used for this purpose, but it is very desirable to complement this by
re-observing many of these sources in the coming years, and if possible, add new targets
to the list for improved robustness. The use and re-calibration of old, possibly unpublished 
data should be pursued. In this context, the unique historical value of positional VLBI 
measurements needs to be stressed. As argued in Sect.~4.5, observers 
should ensure that relevant intermediate data and meta-information are preserved 
for optimal future uses of their data.

\begin{acknowledgements}
A coding error affecting the numerical results in the original version of the paper was 
discovered by S.~Lunz (Potsdam). I am deeply grateful for her careful checking and 
perseverance in hunting down the bug. 
I wish to thank S.~Klioner for many stimulating discussions on this and related topics, and 
for a critical reading of the first draft. I also thank the anonymous referee and 
A.~Bombrun,
U.~Bastian,
M.~Biermann,
J.~Casta{\~n}eda,
M.~Davidson,
C.~Fabricius,
E.~Gerlach,
J.~Hern{\'a}ndez,
D.~Hobbs,
R.~Le Poole,
P.~McMillan,
M.~Ramos-Lerate, and
N.~Rowell
for valuable comments and advice.
Support from the Swedish National Space Board is gratefully acknowledged.
This work has made use of data from the European Space Agency (ESA) mission
{\it Gaia} (\url{https://www.cosmos.esa.int/gaia}), processed by the {\it Gaia}
Data Processing and Analysis Consortium (DPAC,
\url{https://www.cosmos.esa.int/web/gaia/dpac/consortium}). Funding for the DPAC
has been provided by national institutions, in particular the institutions
participating in the {\it Gaia} Multilateral Agreement.
The work has also made use of the SIMBAD database, operated at CDS, Strasbourg, 
France \citep{2000A&AS..143....9W}.
Calculations were made using MATLAB by The MathWorks, Inc.
Diagrams were produced using the astronomy-oriented data handling and visualisation 
software TOPCAT \citepads{2005ASPC..347...29T}.
\end{acknowledgements}

\bibliographystyle{aa} 
\bibliography{refs} 

\appendix

\section{Standard model of stellar motion}\label{sec:stand}

The propagation of stellar astrometric data in time is normally based on a set 
of approximations, assumed to be valid for the observed motion of a single star, or the centre of 
mass of a binary or multiple system, over a limited interval of decades to centuries. Referred to 
in \textit{The Hipparcos and Tycho Catalogues} \citep[][Vol.~1, Sect.~1.2.8]{hip:catalogue} as 
``The Standard Model of Stellar Motion'', these approximations have been adopted as the basis 
for astrometric reductions at least since the days of \citet{1917AJ.....30..137S} and are still used 
for the analysis of \textit{Gaia} data \citep{agis2012}. Among the most important explicit or implicit 
assumptions of the model are (i) that the source moves with constant velocity vector relative 
to the Solar System barycentre, (ii) that light-time effects beyond the Solar System can be ignored, 
and (iii) that aberration effects caused by the curvature of the Galactic orbits can be ignored.
This is not the place to discuss the validity of these assumptions,%
\footnote{Assumptions (i) and (ii) are discussed by \citet{2014A&A...570A..62B} and (iii)
is discussed by \citet{2003A&A...404..743K}. For the practical definition of the astrometric parameters in a 
relativistic framework, see \citet{klioner2003}.}
but because the model, formally represented by the function $\vec{F}_i(\vec{y}_i)$ in 
(\ref{est04}), is central to the paper, it may be useful to summarise the main steps of the
calculation. The Barycentric Celestial Reference System (BCRS) is used throughout, with
times expressed in seconds or Julian years of barycentric coordinate time (TCB), and 
distances in km or au.

\subsection{Propagation of the astrometric parameters}\label{sec:prop}

We first consider the propagation of the astrometric parameters from epoch $T$ to $t$.
Let $\alpha$, $\delta$, $\varpi$, $\mu_{\alpha*}$, $\mu_{\delta}$, and $v_{r}$ 
be the astrometric parameters and radial velocity at the original epoch $T$, and $\alpha(t)$, 
etc.\ their values at epoch $t$. The first step is to 
compute the barycentric unit vector towards the source at time $T$,
\begin{equation}\label{eq:A01}
\vec{r}=\begin{bmatrix} \cos\alpha\cos\delta\\ \sin\alpha\cos\delta\\ 
\sin\delta\end{bmatrix} \, ,
\end{equation}
and the unit vectors in the directions of increasing $\alpha$ and $\delta$,
\begin{equation}\label{eq:A02}
\vec{p}=\begin{bmatrix} -\sin\alpha\\ \cos\alpha\\ 0\end{bmatrix} \, ,\quad
\vec{q}=\begin{bmatrix} -\cos\alpha\sin\delta\\ -\sin\alpha\sin\delta\\ 
\cos\delta\end{bmatrix} \, .
\end{equation}
Next we compute
\begin{equation}\label{eq:A03}
\vec{m} = \mu_{\alpha*}\vec{p}+\mu_{\delta}\vec{q}+(v_{r}\varpi/A)\vec{r} \, ,
\end{equation}
where
\begin{equation}\label{eq:A04}
A=\frac{149\,597\,870.7~\text{km}}{365.25\times 86400~\text{s~yr}^{-1}}  
\simeq 4.740\,470\,464~\text{km~yr~s}^{-1}
\end{equation}
is the astronomical unit. The vectors $\vec{r}$ and $\vec{m}$ are such that
for constant space velocity, the barycentric vector to the source at time $t$ is 
proportional to  
\begin{equation}\label{eq:A05}
\vec{s}(t) = \vec{r} + (t-T)\vec{m} \, .
\end{equation}
The astrometric parameters at epoch $t$ are thus recovered from $\vec{s}(t)$ and $\vec{m}$ 
after rescaling and performing the inverse operations of (\ref{eq:A01})--(\ref{eq:A03}):
\begin{align}
\vec{r}(t) &= |\vec{s}(t)|^{-1}\vec{s}(t)\, , \quad
\vec{m}(t) =|\vec{s}(t)|^{-1}\vec{m}\, , \label{eq:A06}\\
\alpha(t) &= \text{atan2}\Bigl(r_Y(t), r_X(t)\Bigr)\, ,\label{eq:A07}\\ 
\delta(t) &=\text{atan2}\left(r_Z(t),\sqrt{r_X(t)^2+r_Y(t)^2}\,\right)\, , \label{eq:A08}\\
\varpi(t) &= |\vec{s}(t)|^{-1}\varpi \, ,\label{eq:A09}
\end{align}
\begin{align}
\mu_{\alpha*}(t) &= \vec{p}(t)'\vec{m}(t)\, ,\label{eq:A10}\\
\mu_{\delta}(t) &= \vec{q}(t)'\vec{m}(t)\, ,\label{eq:A11}\\
v_r(t) &= \vec{r}(t)'\vec{m}(t)\,A/\varpi(t)\, ,\label{eq:A12}
\end{align}
where
\begin{equation}\label{eq:A13}
\vec{p}(t)=\begin{bmatrix} -\sin\alpha(t)\\ \cos\alpha(t)\\ 0\end{bmatrix} \, ,\quad
\vec{q}(t)=\begin{bmatrix} -\cos\alpha(t)\sin\delta(t)\\ -\sin\alpha(t)\sin\delta(t)\\ 
\cos\delta(t)\end{bmatrix} \, 
\end{equation}
are the updated vectors along $+\alpha$ and $+\delta$.

\subsection{Coordinate direction to the source}\label{sec:ut}

We now turn to the calculation of the position of the source as obtained from a single 
VLBI measurement at time $t$. As opposed to the barycentric direction $\vec{r}(t)$ discussed
above, we now require the topocentric direction from the observer towards the source. 
In VLBI astrometry, geometric delays are calculated from the source positions and station 
coordinates expressed in the BCRS frame and corrected for the gravitational delay caused 
by bodies in the Solar System \citep[for details, see][]{RevModPhys.70.1393}.
The astrometric position measured by VLBI therefore corresponds to the geometric direction 
from the observer towards the target at the time of observation, unaffected by stellar aberration 
and gravitational deflection. This direction, also known as the coordinate direction 
\citep{murray1983}, is here denoted $\vec{\bar{u}}(t)$. 

The modelling of $\vec{\bar{u}}(t)$ in terms of the astrometric parameters is simple because it 
only involves a shift of origin from the Solar System barycentre to the observer. 
With $\vec{b}(t)$ denoting the position of the observer at the time of observation, 
expressed as BCRS coordinates in au, we have
\begin{equation}\label{eq:A14}
\vec{\bar{u}}(t) = \bigl\langle \vec{s}(t_\text{B}) - \varpi\vec{b}(t) \bigr\rangle \, ,
\end{equation}
where angular brackets signify vector normalisation, 
$\langle\vec{a}\rangle=\vec{a}/|\vec{a}|$.
One small complication in Eq.~(\ref{eq:A14}) is that the position of the source should be evaluated
for the barycentric time $t_\text{B}$ obtained by adding the R{\"o}mer delay to the time of observation,
\begin{equation}\label{eq:A15}
t_\text{B}=t+\vec{r}(t)'\vec{b}(t)/c \, ,
\end{equation}
where $c$ is the speed of light. For an observer on the Earth, the R{\"o}mer delay is at most 
about 500~s. Neglecting the delay produces an error equal to the proper motion of the source over this
time interval, which could amount to 0.17~mas for Barnard's star. While the effect is thus negligible for 
most stars, it is always safer to take it into account. On the other hand, it is an acceptable approximation 
to use $\vec{r}$ instead of $\vec{r}(t)$ in (\ref{eq:A15}). (For the position of the observer, $\vec{b}(t)$,
we can normally take the centre of the Earth, which is readily available from standard ephemerides,
although the diurnal parallax of the nearest star is as much as 32~$\mu$as.)
The celestial coordinates of the source are finally obtained from the $X$, $Y$, $Z$ components 
of $\vec{\bar{u}}(t)$ in analogy with Eqs.~(\ref{eq:A07}) and (\ref{eq:A08}).



\clearpage

\section{Origin of the rotation of the bright reference frame of \textit{Gaia} DR2}
\label{sec:phys}

As mentioned in Sect.~\ref{sec:discRot}, the deviating reference frame for bright 
($G\lesssim 13$) sources in \textit{Gaia} DR2 is caused by a deficiency in the
astrometric instrument calibration model used for DR2. Although a detailed discussion 
of the effect is beyond the scope of this paper, it may be useful to outline the basic  
mechanism and how the effect can be avoided in future data releases. Further details
may be found in the documentation of forthcoming releases.
 
The astrometric calibration model used for \textit{Gaia} DR2 is described in Sect.~3.3 of 
\citet{2018A&A...616A...2L}. It contains several terms that depend on the ``window class''
(WC) of the individual CCD observation. Window classes WC0/1/2 are different schemes for 
sampling the pixels around a detected source (for details, see Sect.~3.3.5 in 
\citealt{2016A&A...595A...1G}). The WC is decided by an on-board algorithm, based mainly 
on the $G$ magnitude of the source estimated as it enters the field of view.
WC0 ($G\lesssim 13$) provides a small two-dimensional image of the source, from which 
both the along- and across-scan coordinates can be derived. 
For WC1 ($13\lesssim G\lesssim 16$) and WC2 ($G\gtrsim 16$), the image is marginalised
in the across-scan direction, retaining a one-dimensional array, of different lengths in 
WC1 and WC2, from which only the along-scan coordinate can be derived.

In the cyclic data-processing scheme adopted for \textit{Gaia}
\citep[see Sect.~7.2 of][]{2016A&A...595A...1G}, the astrometric global iterative solution (AGIS) 
is closely connected to a preceding step known as the intermediate data update (IDU).
The IDU provides image centroids to AGIS, which are one- or two-dimensional depending
on the WC; in turn, AGIS furnishes the IDU with the improved astrometry needed to  
calibrate the line- and point-spread functions used by the IDU for centroiding. AGIS and IDU are 
thus inseparable parts of a larger astrometric task, and the calibrations discussed below 
should be understood as including the relevant parts of the IDU calibrations, also done 
per WC.


The use of different calibration models for the three window classes effectively means
that there are three separate instruments to calibrate, one for each WC.%
\footnote{In view of the extreme accuracy goals of \textit{Gaia}, this is an inevitable 
consequence of using different observation modes. The use of TDI blocking gates, 
explained further below, means that there are in fact even more than three separate 
instruments to consider.}
Because the WC
essentially depends on the magnitude of the source, most sources are always observed
in the same WC. Disregarding, for the moment, the sources that are observed in more 
than one WC, we have three disjoint subsets of sources, each subset being used 
to calibrate one WC.
If the calibrations are allowed to be arbitrary functions of time, each subset could have
its own reference frame, with an arbitrary offset in orientation and spin, without causing
any inconsistency in the observations. All that is needed is that the time-dependent 
calibration of each WC at any time exactly matches the positional offsets of the 
sources in the corresponding subset, caused by the rotation of its frame. Although there
is no particular reason why the calibration should vary in exactly this way, it is inevitable
that unrelated variations caused by model imperfections contain some such component.

In reality, there is a good deal of overlap between the window classes, such that many
sources around $G\simeq 13$ are sometimes observed in WC0, and other times in WC1, 
and similarly around $G\simeq 16$ for WC1 and 2. This should in principle suffice 
to guarantee a consistent reference frame across all magnitudes because it is not
possible to obtain consistent solutions for the positions and proper motions of the 
multi-WC sources unless these parameters are expressed in the same reference frame
for all the observations. Because there is no evidence of a discontinuity in spin at $G\simeq 16$
\citep[Fig.~4 in][]{2018A&A...616A...2L}, the overlap indeed seems to have provided a good
connection between the reference frames of WC1 and 2. As shown in the same figure, however, 
it did not work for the WC0/1 transition, or perhaps a more gradual change was created
from $G\simeq 13$ to $\simeq\! 11$ (Sect.~\ref{sec:alt}). 

Of relevance here is that the modelling of observations in WC0 is, for a number of reasons, 
much more difficult than in WC1 and 2. The two-dimensional images in 
WC0 require the point-spread function to be modelled in both the along- and across-scan 
directions, in contrast to the simpler line-spread functions used in WC1 and 2.
The across-scan profile has additional dependences on the precession-induced drift rate, 
resulting in a more complex calibration. 
Furthermore, to increase the dynamic range of \textit{Gaia}'s CCDs, stars brighter than 
$G\simeq 12$ normally obtain reduced integration time through the activation of 
TDI blocking gates \citep{2016A&A...595A...6C}, and the gates also require 
separate calibrations. In spite of the gates, a substantial fraction of the WC0 
observations are affected by pixel saturation, which causes additional issues. 
In \textit{Gaia} DR2 the calibration of WC0 was in fact far from satisfactory, 
as is readily seen in a plot of the RMS post-fit residual versus magnitude 
\citep[Fig.~9 in][]{2018A&A...616A...2L}. This prevented a consistent astrometric 
modelling of the multi-WC sources across the WC0/1 boundary, which may have 
been the direct cause of the observed rotation.

To avoid a similar problem in future \textit{Gaia} data releases, it is necessary
to improve the modelling of WC0 observations. This is already part of ongoing 
activities towards the next releases. It will also help to maximise the
number of multi-WC sources in the primary astrometric solution. Additionally, the 
calibration models for the different window classes may be constrained not to 
contain any time-dependent components representing a 
rotation difference between the corresponding reference frames. Although these 
steps should ensure a consistent reference frame for all window classes, it remains 
important that the consistency can be confirmed, for example by means of VLBI 
observations. 

\end{document}